\documentclass[useAMS,usenatbib,10pt]{mnras}
\usepackage{listings}
\usepackage{times}  
\usepackage{graphics}
\usepackage{subfigure}
\usepackage{amssymb}
\usepackage{aas_macros}
\usepackage{lscape}
\usepackage{hyperref}
\usepackage{rotating}
\usepackage[export]{adjustbox}
\usepackage{graphicx}
\def\gsim{\ga}
\def\lsim{\la}

\def\eagle{{\sc eagle}}
\def\gsim{ \lower .75ex \hbox{$\sim$} \llap{\raise .27ex \hbox{$>$}} }
\def\lsim{ \lower .75ex \hbox{$\sim$} \llap{\raise .27ex \hbox{$<$}} }
\def\simprop{ \lower .75ex \hbox{$\sim$} \llap{\raise .27ex \hbox{$\propto$}} }

\title[The impact of mergers on the AM of galaxies]{
Quantifying the impact of mergers on the angular momentum of simulated galaxies}
\author[Claudia del P. Lagos et al.]{
\parbox[t]{\textwidth}{
\vspace{-1.0cm}
Claudia del P. Lagos$^{1,2}$\thanks{E-mail: claudia.lagos@icrar.org}, Adam R. H. Stevens$^{3}$, Richard G. Bower$^{4}$, Timothy A. Davis$^{5}$, Sergio Contreras$^{6}$, Nelson D. Padilla$^{6,7}$, Danail Obreschkow$^{1,2}$, Darren Croton$^{3}$, James W. Trayford$^{4}$, Charlotte Welker$^{1,2}$, Tom Theuns$^{4}$}
\vspace*{6pt} \\
$^{1}$International Centre for Radio Astronomy Research (ICRAR), M468, University of Western Australia, 35 Stirling Hwy, Crawley, WA 6009, Australia.\\
$^{2}$Australian Research Council Centre of Excellence for All-sky Astrophysics (CAASTRO), 44 Rosehill Street Redfern, NSW 2016, Australia.\\
$^{3}$Centre for Astrophysics \& Supercomputing, Swinburne University of Technology, Hawthorn, VIC 3122, Australia.\\
$^{4}$Institute for Computational Cosmology, Department of Physics, University of Durham, South Road, Durham, DH1 3LE, UK.\\
$^{5}$Astronomy, Cardiff University, Queens Buildings, The Parade, Cardiff CF24 3AA, United Kingdom.\\
$^{6}$Instituto de Astrof\'isica, Pontificia Universidad Cat\'olica de Chile, Avda. Vicu\~na Mackenna 4860, 782-0436 Macul, Santiago, Chile.\\
$^{7}$Centro de Astro-Ingenier\'ia, Pontificia Universidad Cat\'olica de Chile, Avda. Vicu\~na Mackenna 4860, 782-0436 Macul, Santiago, Chile.
\vspace*{-0.5cm}}

\begin{document}


\pagerange{\pageref{firstpage}--\pageref{lastpage}} \pubyear{2016}

\maketitle

\label{firstpage}

\begin{abstract}
We use \eagle\ to quantify the effect galaxy mergers have on the stellar specific angular momentum of galaxies, $j_{\rm stars}$. 
We split mergers into: dry (gas-poor)/wet (gas-rich), major/minor, and different spin alignments and orbital parameters. 
Wet (dry) mergers have an average neutral gas-to-stellar mass ratio of $1.1$ ($0.02$), while major (minor) mergers
are those with stellar mass ratios $\ge 0.3$ ($0.1-0.3$). We correlate the positions of galaxies in the $j_{\rm stars}$-stellar mass plane at $z=0$ with their merger history, and find that galaxies of low spins suffered dry mergers, while galaxies of normal/high spins suffered predominantly wet mergers, if any. The radial $j_{\rm stars}$ profiles of galaxies that went through dry mergers are deficient
	by $\approx 0.3$~dex at $r\lesssim 10\,r_{50}$ (with $r_{50}$ being the half-stellar mass radius), compared to galaxies that went through wet mergers.
Studying the merger remnants reveals that
dry mergers reduce $j_{\rm stars}$ by $\approx 30$\%, while wet mergers increase it by $\approx 10$\%, on average. The latter is connected to the build-up of the bulge by newly formed stars of high rotational speed. Moving from minor to major mergers accentuates these effects. When the spin vectors of the galaxies prior to the dry merger are misaligned,
$j_{\rm stars}$ decreases to a greater magnitude, while in wet mergers 
co-rotation and high orbital angular momentum efficiently spun-up galaxies.
We predict what would be the observational signatures in the $j_{\rm stars}$ profiles driven by dry mergers: (i) shallow radial profiles and (ii) profiles that rise beyond $\approx 10\,r_{50}$, both of which are significantly different from spiral galaxies.
\end{abstract}

\begin{keywords}
galaxies: formation - galaxies: evolution - galaxies: fundamental parameters - galaxies: structure  
\end{keywords}

\section{Introduction}

Galaxy mergers are a natural consequence of the hierarchical growth of structures \citep{White78}
and since early on have been posed to be a key physical process in their morphological transformation 
 (e.g. \citealt{Toomre72,Toomre77,White78b,Farouki82,Barnes88}). Since then, galaxy mergers 
have become an essential process in cosmological galaxy formation models 
(e.g. \citealt{Cole00,Springel01,DeLucia06,Bower06,Lagos08}; see \citealt{Baugh06} for a review).

In the context of the angular momentum (AM) of galaxies, \citet{Fall83} presented the first observational compilation 
of the specific AM of the stellar component of galaxies, $j_{\rm stars}$, and its relation with 
stellar mass, $M_{\rm stars}$. \citet{Fall83} found that elliptical and spiral galaxies follow parallel sequences, with the 
former having $j_{\rm stars}$ a factor of $\approx 6$ lower than the latter. \citet{Fall83} concluded that in hierarchical cosmologies
the $j_{\rm stars}$ values of spirals and ellipticals could be understood if spirals roughly 
conserve $j$ in their formation process (see also \citealt{Mo98}), 
while ellipticals suffer efficient $j$ dissipation.
Galaxy mergers are a natural dissipative phenomenon which could account for the galaxy population of low 
spins.
Early simulations (e.g. \citealt{Barnes87,Navarro94,Heyl96,Zavala08}) showed that dynamical friction can efficiently move 
high $j$ material to the outer regions of galaxies, effectively lowering the $j_{\rm stars}$ of the stellar component that 
is easily measurable. Later on, \citet{Romanowsky12}, via idealised models within the $\Lambda$ cold dark matter ($\Lambda$CDM) paradigm, 
showed that galaxy mergers can naturally explain the positions of elliptical galaxies in the 
$j_{\rm stars}-M_{\rm stars}$ plane, and that
disks and bulges follow fundamentally different $j_{\rm stars}-M_{\rm stars}$ relations.
Recently, using the \eagle\ simulations, 
{\citet{Zavala15} showed that the AM loss of a galaxy's stellar component follows closely that of the inner
parts of its halos, which would be naturally explained by the merging activity of halos and galaxies
at low redshifts.}
Using the same simulations, \citet{Lagos16b} found that mergers were not the only responsible of 
small spins, but that galaxies could also have low $j_{\rm stars}$ due to 
early quenching. 

Recent observational measurements of $j_{\rm stars}$ 
using the Sydney-AAO Multi-object Integral field unit (IFU) spectrograph 
(SAMI; \citealt{Croom12}) by \citet{Cortese16}, have suggested that galaxies form a continuous sequence
in the $j_{\rm stars}-M_{\rm stars}$ plane, instead of the two sequences originally found by \citet{Fall83}. 
\citet{Cortese16} found that the positions of galaxies 
in the $j_{\rm stars}-M_{\rm stars}$ plane were strongly correlated with the Hubble morphological type, S\`ersic index and 
the spin parameter of the stars $\lambda_{\rm R}$, 
which provides a measurement of how rotationally supported a galaxy is \citep{Emsellem07}.
\citet{Cortese16} concluded that the large-scale morphology of galaxies is regulated by their mass and dynamical state.
Similarly, \citet{Obreschkow14b} showed that the relation between the disk $j$ and mass 
has a scatter that correlates with the bulge-to-total mass ratio, arguing that 
the physical processes giving rise to the bulge also affect the formation of the disk, 
and thus there may not be a fundamental distinction between bulges and disks. {It is unclear 
though how much of this result is driven by the sample being dominated by pseudo rather than 
classic bulges.}

This may not, however, be the full story. 
\citet{Emsellem11} showed that early-type galaxies, from the $ATLAS^{\rm 3D}$ survey, 
have a large variety of $\lambda_{\rm R}$ values and thus they cannot be seen as one uniform type of galaxy.
\citet{Emsellem11} found two broad classifications for early-type galaxies: fast and slow rotators.
 Some important trends found by \citet{Emsellem11} and extended recently to higher stellar masses 
by \citet{Veale17}, is that the fraction of slow rotators increases steeply with stellar mass, and that 
the vast majority of S0 galaxies are fast rotators. 
All these observations measure kinematics of galaxies within 
a relatively small area of the galaxy (typically $1$ effective radius), 
which leaves open the question of whether galaxies with low spins 
are the result of a major loss of total $j_{\rm stars}$ or simply a rearrangement of $j_{\rm stars}$ in spite 
of total $j$ conservation. These formation scenarios are not mutually exclusive, and thus 
one has to ask what gives rise to such variety of observed dynamical states
in galaxies, and particularly, early-types.

\citet{Jesseit09}, \citet{Bois11} and \citet{Naab14} found that 
the formation paths of slow and fast rotators can be very varied. For example, \citet{Naab14} showed that slow rotators 
could be formed as a result of wet major mergers, dry major mergers and dry minor mergers. In the case of 
wet mergers, the remnant can be either fast or slow rotators, or even disks (e.g. 
\citealt{Bekki98}; \citealt{Springel00}; \citealt{Cox06}; \citealt{Robertson06}; \citealt{Johansson09}; \citealt{Peirani10}; \citealt{Lotz10};  
\citealt{Naab14}; \citealt{Moreno15}; \citealt{Sparre16a}). 
\citet{DiMatteo09} showed that even dry major mergers of pressure supported galaxies can result in a rotation-supported disk 
if the orbital AM is large enough and efficiently transferred into the orbits of stars.
Many of these mergers may result in a dramatic change in the morphology and spin of galaxies, but ultimately 
mergers are one of many physical processes at play, and continuing gas accretion and star formation can 
reshape the remnant morphology and kinematics. Recently, using cosmological hydrodynamical
simulations, \citet{Sparre16b} found
 that galaxy remnants of major mergers evolve into star-forming disk galaxies unless sufficiently strong feedback 
is present to prevent the disk regrowth. This feedback is an essential mechanism in the new generation of cosmological hydrodynamical simulations, 
such as \eagle\ \citep{Schaye14}, Illustris \citep{Vogelsberger14b} and Horizon-AGN \citep{Dubois14}, 
and most likely plays a major role in reproducing the morphological diversity seen in galaxies \citep{Dubois16, Correa17}. 

Although there is extensive literature for how different merger configurations can affect galaxies, 
cosmological hydrodynamical simulations 
are necessary to realistically represent the frequency of them in a galaxy population, and thus it is the best way of 
shedding light on 
why galaxies display the diversity of $j_{\rm stars}$ seen in observations, especially as modern simulations reproduce the observations 
well \citep{Teklu15,Pedrosa15,Genel15,Zavala15,Lagos16b,Sokolowska16}.
This is the motivation of this work. We use the \eagle\ \citep{Schaye14} cosmological hydrodynamical simulations to 
statistically study how galaxy mergers drive the positions of galaxies in the $j_{\rm stars}-M_{\rm stars}$ plane. 
We also study the main parameters determining how much spin-up or down occurs, and the cumulative effect mergers may have 
in the $z=0$ galaxy population. 
\eagle\ has a good compromise between volume, $(100\,\rm Mpc)^3$, and resolution, $700$~pc, that allows to us have a statistically significant 
sample of galaxies (several thousands with $M_{\rm star}>10^9\,\rm M_{\odot}$) and with enough structural detail to be able to study
mean radial $j_{\rm stars}$ profiles. 

\eagle\ has now been extensively tested against local and high-redshift observations of galaxy sizes \citep{Furlong15}, colours \citep{Trayford15}, 
stellar masses and star formation rates (SFRs; \citealt{Schaye14}; \citealt{Furlong14}), and the gas contents of galaxies \citep{Bahe15,Lagos15,Lagos15b,Crain16}, among 
other properties, with high success. This gives us some confidence 
that we can use \eagle\ to learn about the role of galaxy mergers in the 
$j_{\rm stars}-M_{\rm stars}$ plane. The advent of IFU surveys, such as SAMI, MaNGA \citep{Bundy15} and MUSE \citep{Bacon10}, and the first global measurements of $j_{\rm stars}$ 
at high redshift (\citealt{Burkert16}; \citealt{Swinbank17}; \citealt{Harrison17}), make our study very timely.
{Previous work studying AM in \eagle\ has focused on the galaxy/halo connection \citep{Zavala15}, 
the star formation history effect on the AM \citep{Lagos16b} and 
the structure of the AM and cooling gas in star-forming galaxies \citet{Stevens16b}. In this paper, we therefore 
present an independent, but complementary study of AM in galaxies.}

This paper is organised as follows. $\S$~\ref{EagleSec} briefly describes the \eagle\ simulation and 
introduces the parameters of mergers we study. Here we also present a comparison with observational measurements of 
merger rates, to show that the frequency of mergers is well represented in \eagle. 
In $\S$~\ref{MechanismsInJ} we study the cumulative effect of galaxy mergers as seen by dissecting the $z=0$ galaxy population. We also 
compare the mean radial $j_{\rm stars}$ profiles in \eagle\ with observations of early-type galaxies.
 We then focus on the effect galaxy mergers have on the net value of $j_{\rm stars}$ as well as its radial distribution in galaxies, 
splitting mergers into minor/major, wet/dry and in spin and orbital parameters.
Here we also connect the change in $j_{\rm stars}$ with changes 
in the stellar mass distribution, and analyse the distribution of the stellar components of the galaxies prior to the merger 
 and in the remnant. We present a discussion of our results and our main conclusions in $\S$~\ref{conclusions}.
Finally, in Appendix~\ref{ConvTests} we present a convergence study to show that $j_{\rm stars}$ is well converged 
for the purpose of our study, in Appendix~\ref{SnipshotTests} we analyse the robustness of our result against the 
time resolution of the main simulation used here, 
while Appendix~\ref{ExtraTests} presents additional plots that aid the interpretation of our results.

\section{The EAGLE simulation}\label{EagleSec}

The \eagle\ simulation suite (\citealt{Schaye14}, hereafter
S15, and \citealt{Crain15}, hereafter C15) consists of a large number of cosmological
hydrodynamic simulations with different resolutions, volumes and subgrid models,
adopting the \citet{Planck14} cosmology.
S15 introduced a reference model, within which the parameters of the
sub-grid models governing energy feedback from stars and accreting black holes (BHs) were calibrated to ensure a
good match to the $z=0.1$ galaxy stellar mass function, 
the sizes of present-day disk galaxies and the black hole-stellar mass relation (see C15 for details).
\begin{table}
\begin{center}
  \caption{Features of the Ref-L100N1504 \eagle\ simulation used in this paper. The row list:
    (1) comoving box size, (2) number
    of particles, (3) initial particle masses of gas and (4) dark
    matter, (5) comoving gravitational
    softening length, and (6) maximum physical comoving Plummer-equivalent
    gravitational softening length. Units are indicated in each row. \eagle\
    adopts (5) as the softening length at $z\ge 2.8$, and (6) at $z<2.8$. }\label{TableSimus}
\begin{tabular}{l l l l}
\\[3pt]
\hline
& Property & Units & Value \\
\hline
(1)& $L$ & $[\rm cMpc]$ & $100$\\
(2)& \# particles &  & $2\times 1504^3$ \\
(3)& gas particle mass & $[\rm M_{\odot}]$ & $1.81\times 10^6$\\
(4)& DM particle mass & $[\rm M_{\odot}]$ & $9.7\times 10^6$\\
(5)& Softening length & $[\rm ckpc]$ & $2.66$\\
(6)& max. gravitational softening & $[\rm pkpc]$& $0.7$ \\
\hline
\end{tabular}
\end{center}
\end{table}

In Table~\ref{TableSimus} we summarise the parameters
of the simulation used in this work.
Throughout the text we use pkpc to denote proper kiloparsecs and
cMpc to denote comoving megaparsecs.
A major aspect of the \eagle\ project is the use of
state-of-the-art sub-grid models that capture unresolved physics.
The sub-grid physics modules adopted by \eagle\ are: (i) radiative cooling and
photoheating, (ii) star formation, (iii) stellar evolution and enrichment,
(iv) stellar feedback, and (v) black hole growth and active galactic nucleus (AGN) feedback.
In addition,
the fraction of atomic and molecular gas in each gas particle is calculated in post-processing
following \citet{Lagos15}.

The \eagle\ simulations were performed using an extensively modified
version of the parallel $N$-body smoothed particle hydrodynamics (SPH)
code {\sc gadget-3} (\citealt{Springel08}; \citealt{Springel05b}).
Among those modifications are updates to the SPH technique, which are collectively referred to as
`Anarchy' (see \citealt{Schaller15b} for an analysis of the impact of these changes on
the properties of simulated galaxies compared to standard SPH). 
We use {\sc SUBFIND}
 (\citealt{Springel01}; \citealt{Dolag09}) to identify self-bound overdensities of particles within halos (i.e. substructures).
These substructures are the galaxies in \eagle.

\subsection{Merger parameters studied}\label{mergerpars}

We identify mergers using the merger trees available in the \eagle\ database \citep{McAlpine15}.
These merger trees were created using the $\tt D-Trees$ algorithm of \citet{Jiang14}. \citet{Qu17}
described how this algorithm was adapted to work with \eagle\ outputs. 
Galaxies that went through mergers have more than one progenitor, and for our purpose, 
we track the most massive progenitors of merged galaxies, and compare the properties 
of those with that of the merger remnant to analyse the effect on $j_{\rm stars}$.
The trees stored in the public database of \eagle\ connect $29$ epochs.
The time span between snapshots can range from $\approx 0.3$~Gyr to $\approx 1$~Gyr.
We use these snapshots to analyse the evolution of $j_{\rm stars}$ in galaxies 
and the 
effect of mergers. We consider the interval between outputs appropriate, as our purpose is to analyse 
galaxies before and after, rather than during the merger. 
We study the robustness of our analysis to the time interval between outputs used in the 
simulations in Appendix~\ref{SnipshotTests} using 
much finer time intervals (i.e. snipshots; S15). We find that our calculations 
 are robust and do not sensitively depend on how fine the time interval between outputs are. 
 
We split mergers into major and minor mergers.
The former are those with a stellar mass ratio between the secondary and the primary galaxy $\ge 0.3$, 
while minor mergers have a mass ratio between $0.1$ and $0.3$. Lower mass ratios are considered unresolved and thus 
are classified as accretion \citep{Crain16}.  

In addition to defining minor and major mergers, we estimate the ratio of gas to stellar mass of the merger with the aim of 
 classifying them as gas-rich (wet) or gas-poor (dry) mergers. This ratio is defined as:

\begin{equation}
f_{\rm gas,merger} \equiv \frac{M^{\rm s}_{\rm neutral}+M^{\rm p}_{\rm neutral}}{M^{\rm s}_{\rm stars}+M^{\rm p}_{\rm stars}},
\label{fgasmerg}
\end{equation} 

\noindent where $M^{\rm s}_{\rm neutral}$ and $M^{\rm p}_{\rm neutral}$ are the neutral gas masses of the secondary and primary galaxies, 
respectively,  
while $M^{\rm s}_{\rm stars}$ and $M^{\rm p}_{\rm stars}$ are the corresponding stellar masses.
Masses are measured within an aperture of $30$~pkpc. Neutral gas fractions of individual 
particles in \eagle\ are calculated as 
in \citet{Lagos15}. 
Here, neutral gas refers to atomic plus molecular gas.

Fig.~\ref{FrequencyMergersGasRich} shows the distribution of $f_{\rm gas,merger}$ in three redshift bins in \eagle. 
We find that the distributions are mostly bimodal, and we use this to define gas-rich ($f_{\rm gas,merger}\ge 0.5$) 
and gas-poor ($f_{\rm gas,merger}\le 0.2$) mergers, as 
shown by the vertical dotted lines. From now on, we name these two sub-samples as `wet' and `dry' mergers, respectively.
At $0\le z\le 3$, these two samples are made of $2,677$ and $1,775$ mergers, respectively, and have 
median $f_{\rm gas,merger}$ of $1.1$ and $0.02$, respectively. 
In the literature, `dry' mergers usually refer to galaxies completely devoid 
of gas (e.g. \citealt{Makino97}; \citealt{Naab06}; \citealt{Taranu13}). However,
the reason behind that definition was mostly technical: mergers were studied with collisionless simulations. 
However, in reality galaxies are expected to have some gas, even in the regime of `red and dead' passive galaxies, as shown 
by ATLAS$^{\rm 3D}$ \citep{Young11,Serra12}. 
\eagle\ allows us to define `dry' mergers in a more realistic fashion, by simply imposing them to be gas poor.
When we dissect $f_{\rm gas,merger}$ into the contributions from the primary (the one with the highest stellar mass) and secondary 
galaxies, we find that at any redshift the total gas fraction is dominated by the primary galaxy. 
In \eagle\ we find a good correlation between $f_{\rm gas,merger}$ of the primary and the secondary galaxy, which is stronger at high redshift. 
This correlation is a 
consequence of the `conformity' of the galaxy population 
(i.e. gas-rich galaxies tend to be surrounded by gas-rich galaxies; \citealt{Kauffmann13}; 
\citealt{Wang15}; \citealt{Hearin16}).

We calculate radial $j$ profiles as in \citet{Lagos16b}, which is a measurement if 
a mass-weighted average $j_{\rm stars}$ within $r$ (i.e. $\equiv |J(<r)|/M(<r)$). We will refer to these measurements as 
`mean radial $j_{\rm stars}$ profiles'.
\citet{Lagos16b} showed that $j_{\rm stars}(r_{50})$, calculated with all 
particles within the half-stellar mass radius $r_{\rm 50}$, 
 converges in \eagle\ at $M_{\rm stars}\gtrsim 10^{9.5}\,\rm M_{\odot}$, and thus we limit 
our sample only to galaxies with stellar masses above that threshold.
\citet{Schaller15} showed that the stellar mass 
radial profiles of galaxies in \eagle\ are well converged to scales of at least $\approx 1.5$~pkpc.
{\citet{Schaller15b} analysed the effect of the hydrodynamic scheme 
on galaxy properties and concluded those were minimal compared to the 
effect of the subgrid modelling, showing that any numerical effects affecting the 
AM of galaxies are much less important compared to 
how the baryon physics is modelled (see also \citealt{Scannapieco12} 
and \citealt{Pakmor16}). 
We also tested the convergence of the $j_{\rm stars}$ profiles 
using higher resolution runs and find good convergence down to $\approx 0.5\,\rm r_{50}$ 
(see Appendix~\ref{ConvTests}). 
 Thus, we consider \eagle\ to have an appropriate resolution to perform our study of the effect of 
mergers on $j_{\rm stars}$.

We calculate two angles determining how mergers are oriented: (i) $\theta_{\rm spin}$, is the angle subtended between the 
$\vec{j}_{\rm stars}(\rm tot)$ vectors of the two galaxies that are about to merge, and (ii) $\theta_{\rm orb}$, 
is the angle between $\vec{j}_{\rm stars}(\rm tot)$ of the primary galaxy and the orbital AM vector,

\begin{equation}
\theta_{\rm spin}={\rm acos}\left[\hat{j}^{\rm s}_{\rm stars}(\rm tot)\cdot \hat{j}^{\rm p}_{\rm stars}(\rm tot)\right],
\label{Eqspin}
\end{equation}

\noindent and

\begin{equation}
\theta_{\rm orb}={\rm acos}\left[\hat{j}_{\rm orbital}\cdot \hat{j}^{\rm p}_{\rm stars}(\rm tot)\right],
\end{equation}

\noindent where $\vec{j}^{\rm s}_{\rm stars}(\rm tot)$ and $\vec{j}^{\rm p}_{\rm stars}(\rm tot)$ are the normalized 
$j_{\rm stars}$ vectors of the secondary and primary galaxies, respectively, and $\vec{j}_{\rm orbital}=\vec{r} \times \vec{v}$. 
Here, $\vec{r}$ and $\vec{v}$ are the position and velocity vectors of the secondary galaxy in the rest-frame of the 
primary one, calculated in the last snapshot the two galaxies were identified as separate objects.
Galaxy growth produced by gas accretion and star formation 
will be termed `smooth accretion' during the rest of the paper.

\begin{figure}
\begin{center}
\includegraphics[trim=2mm 3mm 0mm 2mm, clip,width=0.5\textwidth]{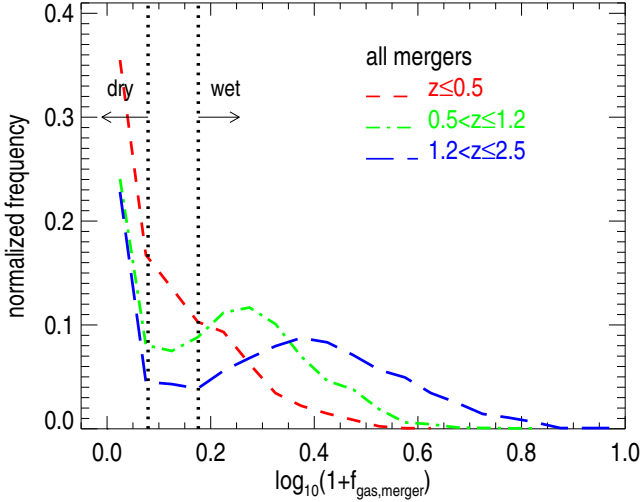}
\caption{The distribution of the neutral gas-to-stellar mass ratio of mergers in \eagle\ in $3$ redshift bins, as labelled. 
Distributions are mostly bimodal, and we use this to define gas-rich (wet) and gas-poor (dry) mergers in \eagle\ (shown as dotted lines).}
\label{FrequencyMergersGasRich}
\end{center}
\end{figure}

\begin{figure}
\begin{center}
\includegraphics[trim=0mm 5mm 0mm 0mm, clip,width=0.45\textwidth]{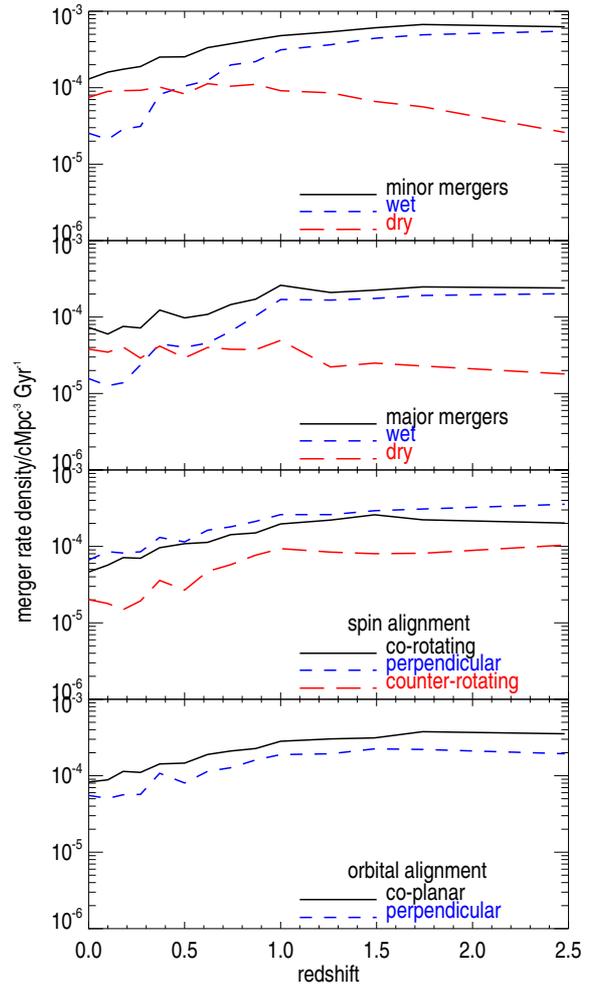}
\caption{Merger rate density as a function of redshift in \eagle.
The top panel shows minor mergers and the subsamples of wet and dry minor mergers, as labelled. 
The middle panel is the same but for major mergers, while the bottom panels show mergers split into 
 spin (i.e. co-rotating, perpendicular and counter-rotating) and orbital (i.e. co-planar and perpendicular) alignments, respectively.}
\label{FrequencyMergers}
\end{center}
\end{figure}

\begin{figure}
\begin{center}
\includegraphics[trim=0mm 10mm 0mm 0mm, clip,width=0.45\textwidth]{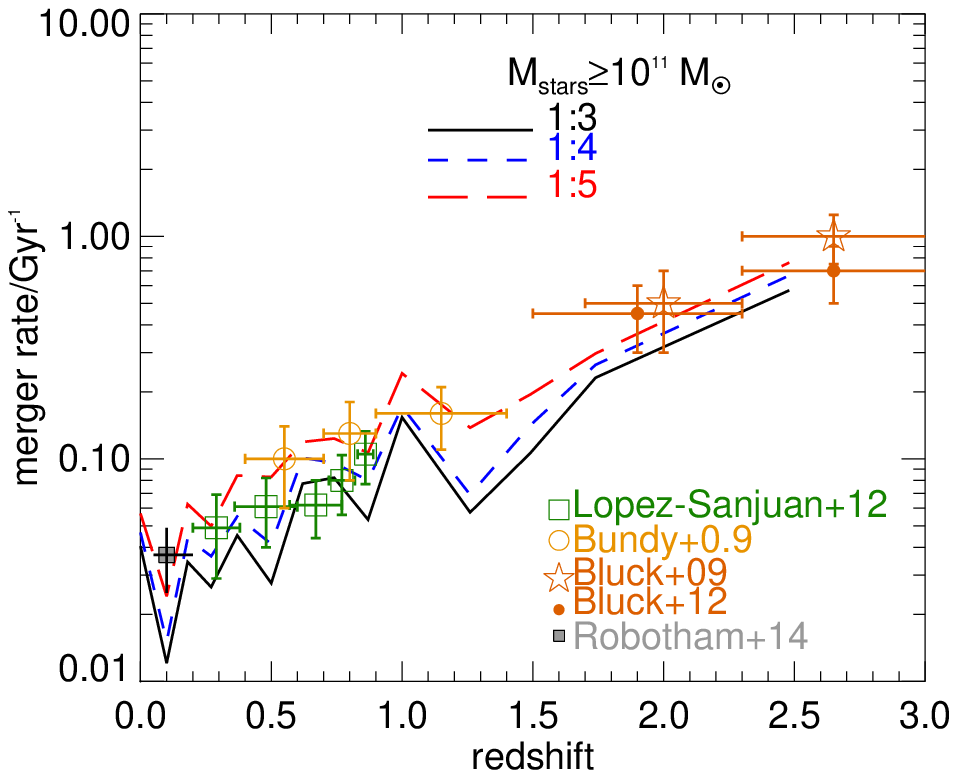}
\caption{Merger rate in galaxies with $M_{\rm stars}\ge 10^{11}\,\rm M_{\odot}$ in \eagle\ 
as a function of redshift. We show merger ratios $\ge 1:3$ (solid line), 
$\ge 1:4$ (short-dashed line) and $\ge 1:5$ (long-dashed line). 
We also show the observational measurements of 
\citet{Bundy09}, \citet{Bluck09}, \citet{Lopez-Sanjuan12}, \citet{Bluck12} and \citet{Robotham14}. 
Most of these observational works assume major mergers are those with stellar mass ratios $\ge 1:4$.
\eagle\ predicts major merger rates of massive galaxies that are in excellent agreement with the 
observations in the entire redshift range where measurements are available.}
\label{FrequencyMergersMassive}
\end{center}
\end{figure}

The top and middle panels of Fig.~\ref{FrequencyMergers} show the merger rate density of minor and major mergers in (primary) galaxies with 
$M_{\rm stars}>10^{9.5}\,\rm M_{\odot}$ as a function of redshift, respectively, and split into 
wet and dry.
The frequency of mergers is noisy due to the small volume of the simulation 
and the relatively high stellar mass threshold we are applying to our 
galaxy sample. The frequency of dry mergers increases from $z=2.5$ down to 
$z=1$ in both minor and major mergers, with an approximately constant frequency at $z<1$. 
The frequency of wet mergers instead decreases steadily towards $z=0$. 
This is driven by \eagle\ galaxies having $f_{\rm gas,merger}$ that decrease with time \citep{Lagos15,Lagos15b}. 
The bottom panels of Fig.~\ref{FrequencyMergers} show the frequency of mergers split by their spin orientation 
and orbital alignment. 
In the case of spin alignments, we define co-rotating, 
perpendicular and counter-rotating mergers as those with $\rm cos(\theta_{\rm spin})>0.7$ (angles between $0-45$~degrees), 
$-0.15<\rm cos(\theta_{\rm spin})<0.15$ (angles between $81-99$~degrees) 
and $\rm cos(\theta_{\rm spin})<-0.7$ (angles between $135-180$~degrees), respectively.
Randomly oriented mergers in three dimensions would imply a uniform distribution in $\rm cos(\theta_{\rm spin})$; 
hence the number of mergers in these equal ranges 
($0.3$ in $\rm cos(\theta_{\rm spin}$) directly show their relative frequency.
We find in \eagle\ that perpendicular mergers are $\approx 1.3$ times more common than co-rotating mergers, 
but counter-rotating mergers are $\approx 3.4$ and $\approx 2.6$ times less common than 
perpendicular and co-rotating mergers, respectively.
In the case of orbital alignments, we define co-planar mergers as those with $\rm |cos(\theta_{\rm orb})|\ge 0.7$, 
while perpendicular mergers have $\rm |cos(\theta_{\rm orb})|\le 0.3$. 
We find that co-planar mergers are $\approx 1.5$ more frequent than perpendicular ones. 
The trends we see here are consistent with those presented 
by \citet{Welker15} using the {\tt Horizon-AGN} simulation. 
\citet{Welker15} showed that satellite galaxies on a decaying orbit towards the central galaxy tend to align with the galactic plane 
of the central in a way that, by the time they merge, are most likely to come in an orbit aligned with the galactic plane of the primary. 
\citet{Welker15} also found that mergers taking place in filaments are more likely to be of galaxies with $\rm cos(\theta_{\rm spin})\approx 0$ 
if the primary galaxy is a passive, spheroidal galaxy, while co-rotation is expected if the primary galaxy is a spiral, star-forming galaxy. 
The frequencies we report in the bottom panels of Fig.~\ref{FrequencyMergers} are consistent with this picture.

Fig.~\ref{FrequencyMergersMassive} compares the major merger 
rate of \eagle\ galaxies with $M_{\rm stars}\ge 10^{11}\,\rm M_{\odot}$ at different redshifts 
against a 
compilation of observations. Here we employ $3$ different stellar mass ratios to define 
major mergers: $\ge 1:5$, $\ge 1:4$ and $\ge 1:3$, to show the systematic variations produced by this definition. 
The observations shown in Fig.~\ref{FrequencyMergersMassive} correspond to measurements coming from 
the characterisation of pair frequency \citep{Bundy09,Bluck09,Lopez-Sanjuan12,Robotham14}, and from 
the identification of galaxies with disturbed morphologies \citep{Bluck12}. Both set of measurements agree remarkably well.
We find that the major merger rate of massive galaxies is in excellent agreement with the observations. 
For our purpose this means that the effect of galaxy mergers on the AM of galaxies is not going to be over(under)-represented.

\section{The effect of mergers on the stellar specific AM of galaxies throughout cosmic time}\label{MechanismsInJ}

In $\S$~\ref{netz0}, we present an analysis of the accumulated effect of mergers by studying the galaxy population 
at $z=0$. In $\S$~\ref{temporalevo} we analyse the effect of mergers by comparing measurements of galaxy properties before and after 
the mergers, and how these depend on the nature of the merger. In $\S$~\ref{rearr-Sec} we analyse the radial rearrangement 
of $j_{\rm stars}$ as a result of mergers.

\begin{figure*}
\begin{center}
\includegraphics[trim=0mm 3mm 0mm -3mm, clip,width=0.9\textwidth]{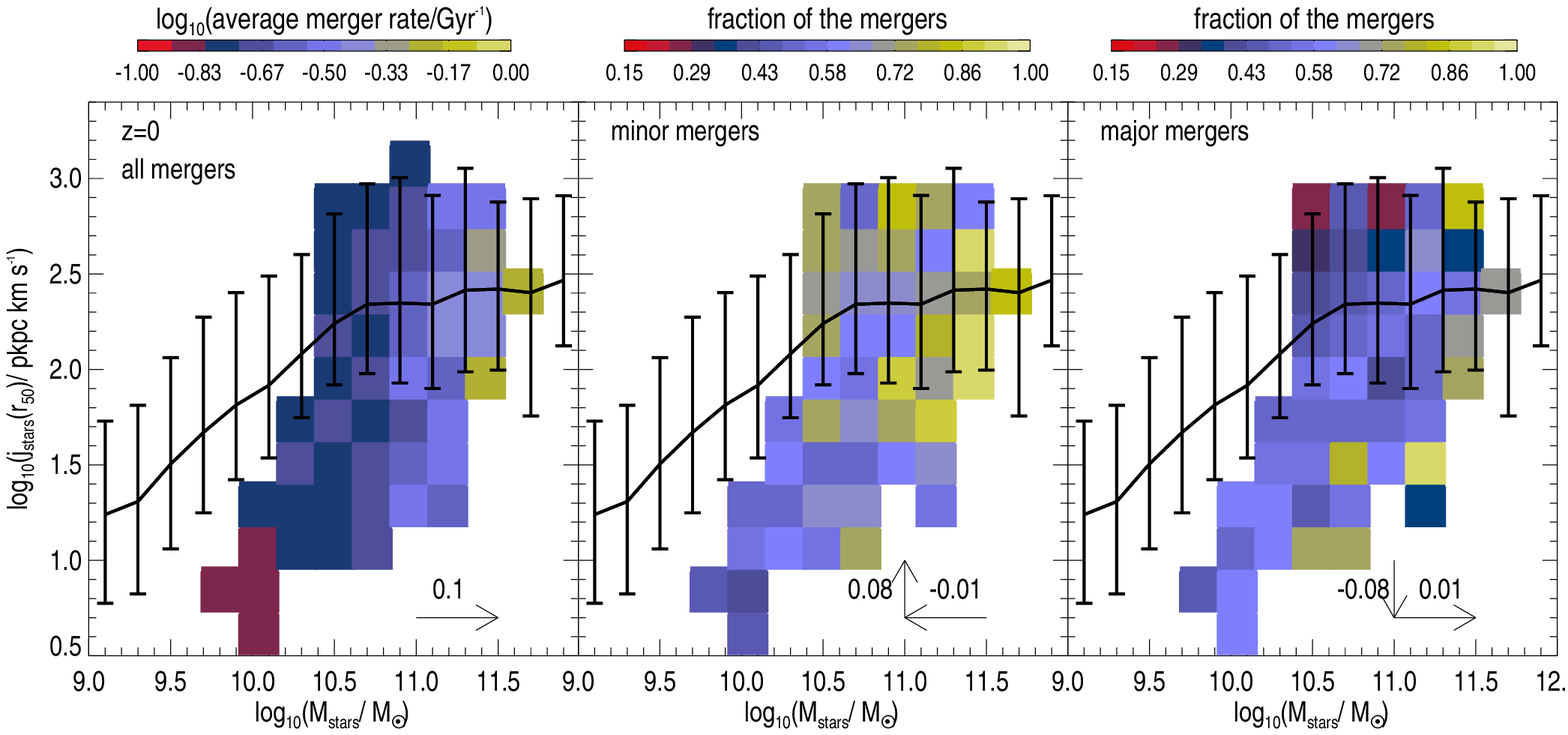}
\includegraphics[trim=0mm 3mm 0mm -3mm, clip,width=0.9\textwidth]{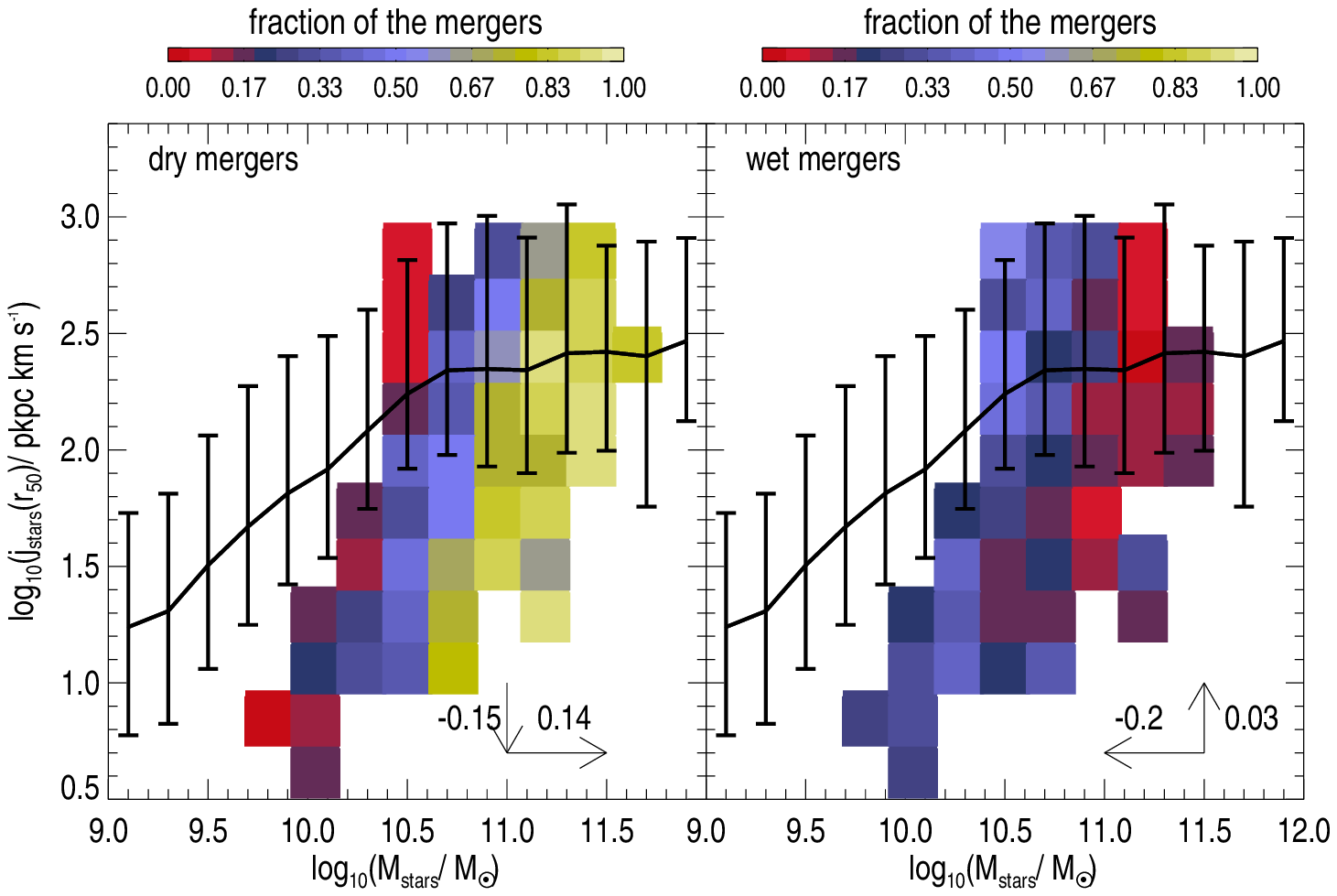}
\caption{{\it Top panels:} The specific AM of the stars measured with all the particles within
the half-mass radius of the stellar component as a function
of stellar mass at $z=0$ for galaxies with $M_{\rm stellar}>10^9\,\rm M_{\odot}$ in \eagle. 
The lines with errorbars show the median and the $16^{\rm th}$ to $84^{\rm th}$ percentile ranges.
In the left panel we colour bins (with $\ge 5$ galaxies) in which more than $50$\% of the galaxies have suffered mergers, 
by the average merger rate per galaxy they had during their lifetimes. 
In the middle and right panels we colored those same bins by the fraction of the mergers that were minor and major, respectively.
By construction, the fractions of the middle and right panels in each $2$-dimensional bin add up to $1$.
{\it Bottom panels:} As in the top middle and right panels, but for the fraction of the mergers that were dry and wet, respectively. 
The arrows in each panel indicate the directions in which the frequency of the respective merger type increases and the number 
next to the arrows show the best fit power-law index for the relations:  
average merger rate $\propto M^{\alpha}$, 
merger fraction $\propto M^{\alpha}$ and merger fraction $\propto j^{\alpha}$.}
\label{Sigmaz0MergersMinorAndMajor}
\end{center}
\end{figure*}

\begin{figure}
\begin{center}
\includegraphics[trim=0mm 0mm 0mm 1mm, clip,width=0.499\textwidth]{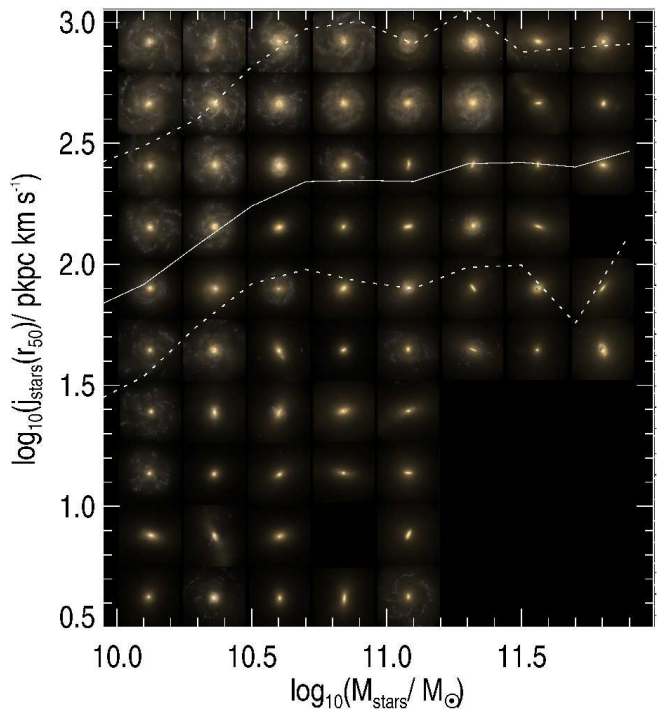}
\caption{Visualisation of the optical morphology of galaxies in the $j_{\rm stars}(r_{50})$-$M_{\rm stars}$ plane at $z=0$. We randomly select 
galaxies 
in $10$ bins of ${\rm log}_{\rm 10}(j_{\rm stars}(r_{50}))$ and $8$ bins of ${\rm log}_{\rm 10}(M_{\rm stars})$ in the range $10^{0.5}-10^3\,\rm pkpc\,km\,s^{-1}$ and 
$10^{10}-10^{12}\,\rm M_{\odot}$, respectively, and show here their synthetic g, r, i face-on optical images. 
Only bins with $\ge 3$ galaxies are shown here. 
These images are $60$~pkpc on a side are are publicly 
available from the \eagle\ database \citep{McAlpine15}. 
The solid and dotted lines show the median and the $16^{\rm th}$ to $84^{\rm th}$ percentile range.}
\label{VisualJstars}
\end{center}
\end{figure}

\begin{figure}
\begin{center}
\includegraphics[trim=0mm 0mm 0mm 1mm, clip,width=0.47\textwidth]{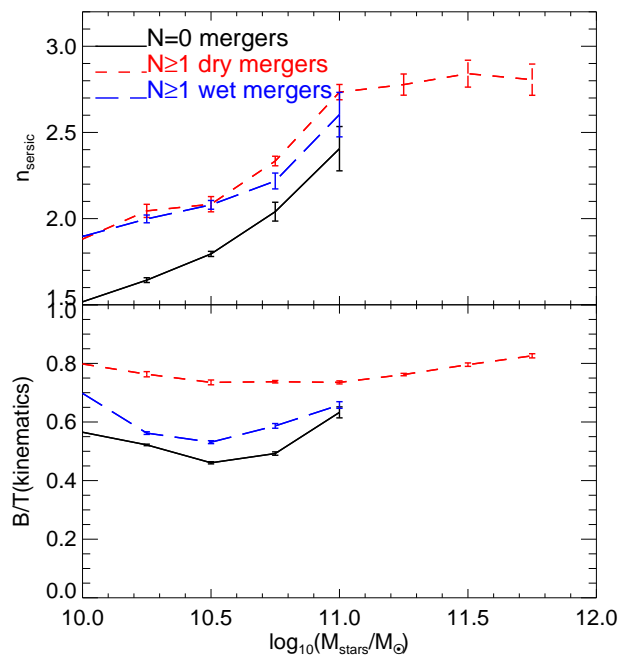}
\caption{{S\'ersic index (top) 
and {kinematic bulge-to-total ratio (bottom)} as a function of stellar mass for galaxies at $z=0$. 
We show galaxies with $M_{\rm stellar}\ge 10^{10}\,\rm M_{\odot}$, which is where galaxy mergers 
become common (see top-left panel of Fig.~\ref{Sigmaz0MergersMinorAndMajor}).
Lines with errorbars show the median and $1\sigma$ error on the median for 
galaxies that have not had mergers, and those that had at least one dry or wet merger, as labelled in the top panel, by $z=0$. 
This figure shows that galaxies that suffered dry mergers have the highest S\'ersic indices and B/T ratios.}}
\label{Nsersic}
\end{center}
\end{figure}

\begin{figure}
\begin{center}
\includegraphics[trim=0mm 4mm 0mm 4mm, clip,width=0.47\textwidth]{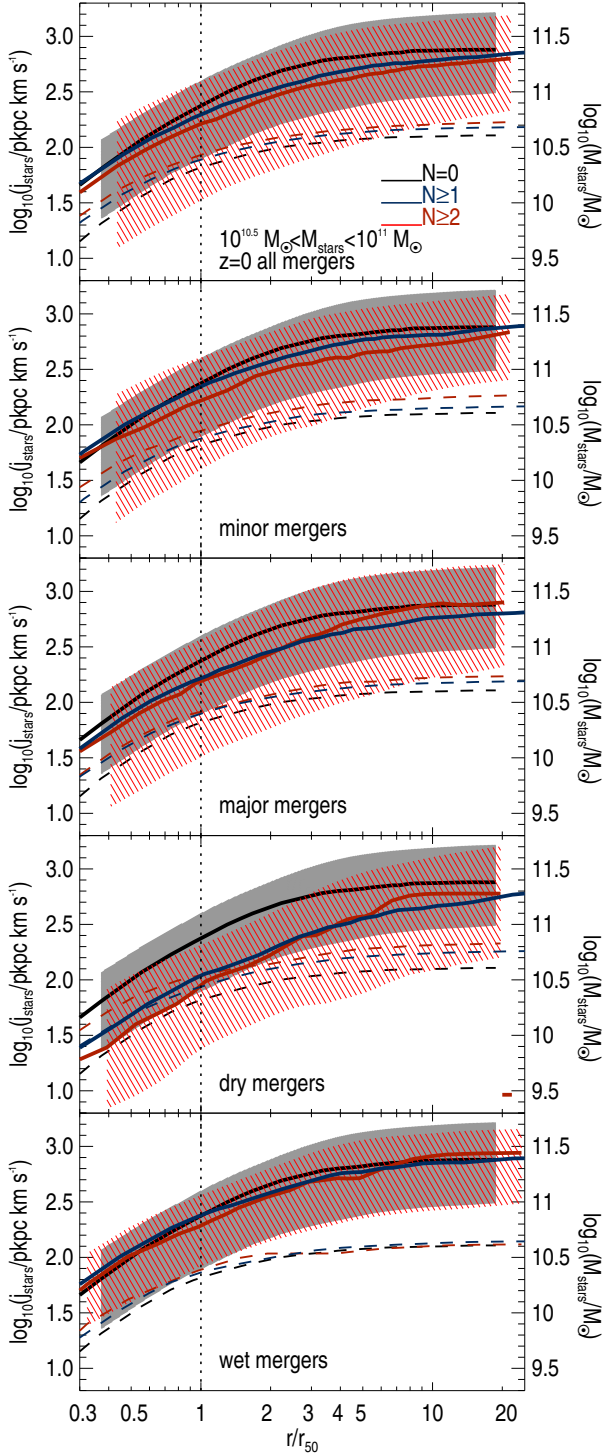}
\caption{{\it Top panel:} $j_{\rm stars}$ (solid lines) and $M_{\rm stars}$ (dashed lines) measured 
within $r$ as a function of $r$ in units of $r_{\rm 50}$ for 
galaxies at $z=0$ with total stellar masses in the range $10^{10.5}\,\rm M_{\odot}-10^{11}\,\rm M_{\odot}$ 
 that have gone through different numbers of galaxy mergers, as labelled. 
Lines show the median of the $j_{\rm stars}$ profiles of galaxies, while the
shaded regions show the $16^{\rm th}$ to $84^{\rm th}$ percentile range, plotted only for 
$N_{\rm mergers}=0,\,\ge 2$, for clarity. 
The scale of $j_{\rm stars}$ and $M_{\rm stars}$ are marked in the left and right axis, respectively. 
 {\it Other panels:} As in the top panel, but distinguishing between minor, major, dry and wet mergers, as labelled.
This figure shows that galaxy mergers generally lead to a deficit of 
$j_{\rm stars}$ at $r\lesssim 10\,r_{50}$, with dry mergers causing pronounced deficits of $\approx 0.5$~dex. 
At sufficiently large radii, $j_{\rm stars}$ converges to a value set by the dark matter halo.}
\label{CumulativeJstars}
\end{center}
\end{figure}

\begin{figure}
\begin{center}
\includegraphics[trim=0mm 4mm 3mm 4mm, clip,width=0.47\textwidth]{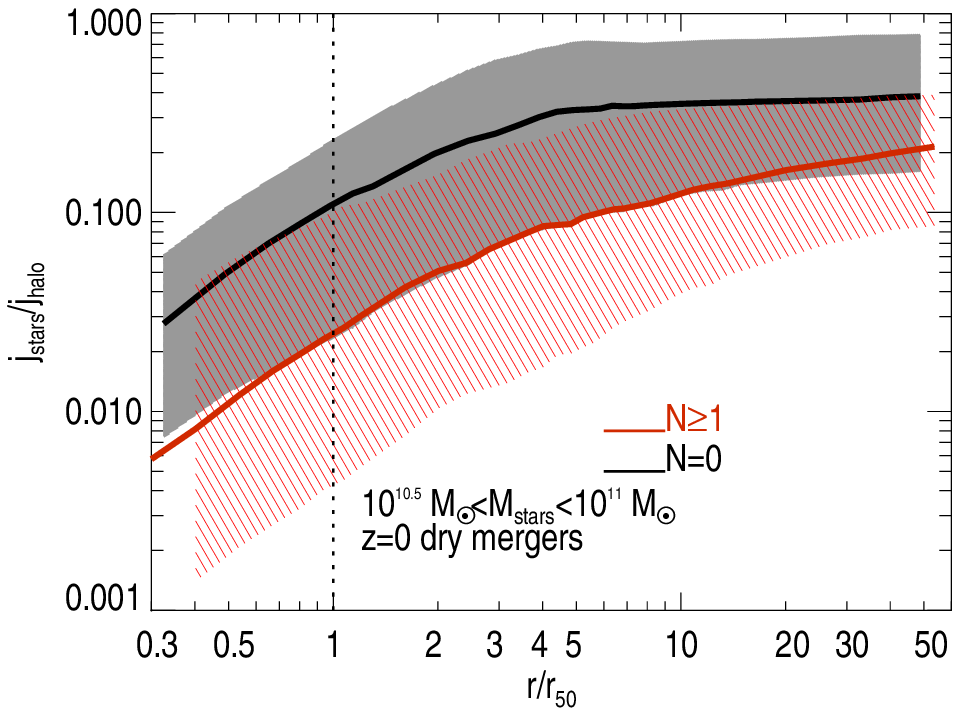}
\caption{As Fig.~\ref{CumulativeJstars}, but for the subsample of central galaxies at $z=0$ that have not suffered a merger 
(black solid line) and 
those that went through at least one dry merger (red solid line), and that have stellar masses 
in the range $10^{10.5}\,\rm M_{\odot}-10^{11}\,\rm M_{\odot}$.
Here we show $j_{\rm stars}$ in units of the specific AM of the dark matter halo, $j_{\rm halo}$, calculated 
with all dark matter particles within the virial radius. Galaxies that did not have a merger typically 
have $j_{\rm stars}$ increasing from $\approx 0.1$ at $r_{\rm 50}$ to $\approx 0.4\,j_{\rm halo}$ at $10\,r_{50}$, while 
galaxies that had at least one dry merger go from $\approx 0.02$ to $\approx 0.2\,j_{\rm halo}$ in the same radii range, on average.}
\label{theorycomp}
\end{center}
\end{figure}

\subsection{The net effect of galaxy mergers seen at $z=0$}\label{netz0}

The top-left panel of Fig.~\ref{Sigmaz0MergersMinorAndMajor} shows how the 
galaxy merger rate changes with the position of galaxies 
in the $j_{\rm stars}(r_{50})$-$M_{\rm stars}$ plane. We define the average merger rate of individual galaxies as the number 
of mergers divided by the stellar-mass weighted 
age. Here we do not distinguish recent from past mergers, but just count their occurrence.
We colour only those bins in which at least $50$\% of the galaxies have undergone mergers during their lifetimes. This is why 
below $M_{\rm stars}\approx 10^{10}\,\rm M_{\odot}$ there are very few coloured bins.
At $10^{10}\,\rm M_{\odot}\lesssim M_{\rm stars}\lesssim 10^{10.5}\,\rm M_{\odot}$ mostly galaxies with low spins 
have a significant contribution from mergers. 
These galaxies are hosted by halos that are on average $20$\% more massive than those of galaxies of the same stellar 
mass but that never had mergers. 
At $M_{\rm stars}\gtrsim 10^{10.5}\,\rm M_{\odot}$ the vast majority of galaxies had at least one merger by $z=0$. 
The merger rate increases with increasing mass (best power-law fit is $\propto M^{0.1}_{\rm stars}$), 
and no clear correlation is seen with $j_{\rm stars}(r_{50})$ at fixed stellar mass. 

In the top middle and right panels of Fig.~\ref{Sigmaz0MergersMinorAndMajor} we calculate the fraction of the mergers shown in the 
left panel that were minor and major, respectively. 
We also performed power-law best fits to the relationship between 
 the merger fraction and $M_{\rm stars}$ and $j_{\rm stars}(r_{50})$ to quantify the trends. 
The best fit power-law indices are shown in each panel of Fig.~\ref{Sigmaz0MergersMinorAndMajor}. 

The fraction of major and minor mergers weakly increase and 
decrease, respectively, with increasing stellar mass (see power-law 
indices in Fig.~\ref{Sigmaz0MergersMinorAndMajor}).
We also see a slightly stronger trend with $j_{\rm stars}(r_{50})$: at fixed stellar mass, the frequency of major and minor mergers 
decrease and increase, respectively with increasing $j_{\rm stars}(r_{50})$.
The directions in which the frequency of mergers increase are shown as arrows in 
Fig.~\ref{Sigmaz0MergersMinorAndMajor}. 

In the bottom panels of Fig.~\ref{Sigmaz0MergersMinorAndMajor} we split the mergers into dry and wet, 
following the definition of Fig.~\ref{FrequencyMergersGasRich}.
We find stronger trends with both $M_{\rm stars}$ and $j_{\rm stars}(r_{50})$.
For dry mergers, 
we find an increase in their frequency with increasing stellar mass, and we identify a significant gradient of 
an increasing fraction of dry mergers with decreasing $j_{\rm stars}(r_{50})$ at fixed stellar mass 
(see power-law               
indices in the bottom panel of Fig.~\ref{Sigmaz0MergersMinorAndMajor}). For wet mergers, we find 
that their fraction increases with decreasing stellar mass and increasing $j_{\rm stars}(r_{50})$. 
The latter though is a very weak trend. 
Fig.~\ref{Sigmaz0MergersMinorAndMajor} indicates that the gas fraction involved 
in the merger is more important than 
the mass ratio for the current $j_{\rm stars}(r_{50})$ of galaxies. We examine 
the same plots for $j_{\rm stars}$ measured within $5\,r_{50}$ (used to encompass the entire galaxy) 
and find the same trends (not shown here). 
These results suggest that galaxy mergers can have a devastating effect on the specific AM of galaxies, but 
with the exact effect strongly depending on the nature of the merger.

\citet{Lagos16b} found that the positions of galaxies in the $j_{\rm stars}(r_{50})$-$M_{\rm stars}$ plane are strongly correlated 
with a galaxy's gas fraction, stellar age, stellar concentration, optical colour and $V/\sigma$, all of which are 
usually used to distinguish early and late type galaxies. In Fig.~\ref{VisualJstars} we explicitly show how the morphology of galaxies 
changes in this plane. Here we randomly selected galaxies in $10$ bins of ${\rm log}_{10}(j_{\rm stars}(r_{50}))$ and 
$8$ bins of ${\rm log}_{\rm 10}(M_{\rm stars})$ in the ranges $10^{0.5}-10^3\,\rm pkpc\,km\,s^{-1}$ and
$10^{10}-10^{12}\,\rm M_{\odot}$, respectively, and extract their synthetic optical images from the \eagle\ database. 
These images are face-on views and are $60$~pkpc on a side. 
This figure shows that at fixed $M_{\rm stars}$, galaxies go from being 
red spheroidals at 
low $j_{\rm stars}(r_{50})$ to being grand-design spirals at high $j_{\rm stars}(r_{50})$ in the stellar mass 
range $10^{10}\,\rm M_{\odot}\lesssim M_{\rm stars}\lesssim 10^{11}\,\rm M_{\odot}$.
At higher stellar masses, galaxies with high $j_{\rm stars}(r_{50})$ appear more like defunct spirals, 
with little star formation and 
aging disks. If we follow the median $j_{\rm stars}(r_{50})$, one sees that galaxies 
go from being preferentially spiral/disk-dominated at 
$M_{\rm stars}\approx 10^{10}\,\rm M_{\odot}$ to spheroids at 
$M_{\rm stars}\gtrsim 10^{11.5}\,\rm M_{\odot}$. Given the strong correlation between the 
positions of galaxies in the $j_{\rm stars}(r_{50})$-$M_{\rm stars}$ plane with the frequency of wet/dry mergers, 
and with the morphologies of galaxies, 
one would expect morphologies to be connected to wet/dry mergers. 
\citet{Rodriguez-Gomez16} showed that the morphologies of galaxies 
are connected with the merger history in the Illustris simulation, 
but they could only determine a clear correlation in massive galaxies, $M_{\rm star}\ge 10^{11}\,\rm M_{\odot}$, due to predominance of 
dry mergers and ex-situ formation of the stars. 
Our results in \eagle\ suggest that the morphology of galaxies, as well as their 
$j_{\rm stars}$, sensitively depend on the type of the merger.

{To corroborate this relation, in Fig.~\ref{Nsersic} we show the $3$-D S\'ersic index (measured from the 
$3$-D stellar mass distributions) and the {kinematic bulge-to-total, B/T, ratio\footnote{{ 
$\rm B/T\equiv 1-\kappa_{co}$, where $\kappa_{\rm co}$ is  
the ratio of kinetic energy invested in ordered rotation calculated using only star particles that follow the direction of rotation
of the galaxy (see \citealt{Correa17} for more details). 
We also analysed the \citet{Abadi03} and \citet{Sales10} definitions of 
kinetic B/T and found the same trends as in the bottom panel of Fig.~\ref{Nsersic}.}}
 as a function of stellar mass}
for galaxies at $z=0$ that have not suffered mergers, and had at least one dry or wet merger. 
This figure shows that no-merger galaxies have much lower S\'ersic indices {and B/T ratios} than galaxies that had mergers. 
Galaxies 
that had dry mergers also have the highest S\'ersic indices and {B/T ratios}, confirming the connection we see between 
the visual morphologies of Fig.~\ref{VisualJstars} and the frequency of dry mergers in 
Fig.~\ref{Sigmaz0MergersMinorAndMajor}. Galaxies that had wet mergers have S\'ersic indices {and B/T ratios} 
lower than the dry merger remnants, but higher 
than the no-merger galaxies. This is consistent with the findings discussed in \S~\ref{WetMersDetails} of the central stellar concentration
 in galaxies increasing during wet mergers.}
We explore the effect of mergers on $j_{\rm stars}(r_{50})$ further in $\S$~\ref{temporalevo}. 

We now examine the mean radial $j_{\rm stars}$ profiles of galaxies at $z=0$ in 
Fig.~\ref{CumulativeJstars} in a narrow range of stellar mass, 
$10^{10.5}\,\rm M_{\odot}\lesssim M_{\rm stars}\lesssim 10^{11}\,\rm M_{\odot}$. In the same Figure we also show the 
cumulative radial profile of 
$M_{\rm stars}$. In the top panel we show how different the radial $j_{\rm stars}$ profiles 
are in galaxies that suffered different numbers of mergers, without 
yet distinguishing the type of merger. Increasing the frequency of mergers has the effect of flattening the radial $j_{\rm stars}$ 
profile, 
in a way that galaxies that went through a higher number of mergers have a deficit 
of $j_{\rm stars}$ at $0.5\,r_{50}\lesssim r\lesssim 10 \,r_{50}$
as large as $\approx 0.3$~dex compared to their counterparts of the same mass but that did not go through mergers. 
At sufficiently large radii, $j_{\rm stars}$ converges so that galaxies with different 
number of mergers have a similar $j_{\rm stars}(\rm tot)$. 
This suggests that the most important effect of mergers is in the radial structure of $j_{\rm stars}$ rather than the 
total $j_{\rm stars}$. The stellar mass cumulative profile of galaxies is also shallower when the 
frequency of mergers increases, which means that 
a larger fraction of the stellar mass is locked up in the outskirts of galaxies.
Although there is a small tendency for galaxies that went through more mergers to have a larger $r_{50}$, 
the trends here are not affected by this, as the differences in the radial profiles 
are very similar even when we do not normalise the $x$-axis by $r_{\rm 50}$. 
By splitting mergers into minor and major (second and third panels of Fig.~\ref{CumulativeJstars}), 
we find that galaxies that had one major merger can have 
a deficit in $j_{\rm stars}$ similar to those that had two minor mergers, 
and increasing the frequency of major mergers does not seem to have the cumulative effect 
that is seen for minor mergers. 
In our sample, \eagle\ galaxies that had major mergers can have 
 minor mergers too, but for the sample of minor mergers, we 
remove all galaxies that had at least one major merger.

We then analyse mergers split into dry and wet in the bottom panels of Fig.~\ref{CumulativeJstars} 
and find that {\it dry mergers have a 
 catastrophic} effect on $j_{\rm stars}$ from the galaxy's centre out to $\approx 20\,r_{50}$. The 
deficit of $j_{\rm stars}$ compared to galaxies without a merger
is as large as $\approx 0.5$~dex. Also note that the stellar mass cumulative profile 
is much shallower for galaxies that went through a dry merger. 
In the case of wet mergers we see the exact opposite. Very little difference is found between 
galaxies that did not suffer a merger, 
and those that suffered 
one, two or more wet mergers. This reinforces the conclusion that to $j_{\rm stars}$ of a galaxy, 
what matters most is whether the merger is dry or wet. 
We will show in Fig.~\ref{HistogramsDeltaJ} that this is also true when we study $j_{\rm stars}$ before and after the merger. 
Note that in the case of dry mergers, we still see that the mean radial $j_{\rm stars}$ 
profile converges at sufficiently large radii 
to a $j_{\rm stars}\rm (tot)$ that does not strongly depend on the merging history of galaxies. 

\subsubsection{The galaxy/halo specific AM connection}

We compare $j_{\rm stars}$ of the galaxies with the specific AM of their dark matter halos in the top panel of 
 Fig.~\ref{theorycomp}. We find that galaxies that went through at least one dry merger on average have a $j_{\rm stars}\rm (tot)$ that is $5$ 
times smaller than that of their halo, while galaxies that did not go through a merger typically retain $\approx 40-50$\% of their halo $j$. 
This latter result agrees very well with the prediction by \citet{Stevens16} for spiral galaxies.  
With a semi-analytic model, those authors evolved the one-dimensional structure of disks in a series of annuli of fixed $j$. 
They assumed that when gas cooled or accreted onto a galaxy, it carried the same total $j$ of the 
halo \emph{at that time} in both magnitude and direction, 
and that is was distributed exponentially. Stars were formed in annuli that were Toomre unstable or had sufficient H$_2$.  
At $z=0$, they found spiral galaxies (which had not suffered dry mergers) had $j_{\rm stars} / j_{\rm halo} = 0.4 \pm 0.29$, independent of galaxy 
mass (see their Fig.~13). Despite the completely different methodology, this aligns almost perfectly with the result of \eagle\ galaxies that have 
not participated in a dry merger.

\citet{Fall83} suggested that spiral galaxies need to have a $j_{\rm stars}$ close to that of their halo 
({within $\approx 80$\% according to \citealt{Fall13}}), 
while ellipticals
 had to lose $90$\% of their $j$, postulating a fundamental difference between the two galaxy populations. 
The conclusions reached by these authors 
were biased by the available observations, that in the best case went out to $\approx 10r_{50}$. 
According to \eagle, early-type galaxies 
only reach $\approx 0.1$ of the expected halo $j$ at $r\approx 10r_{50}$, on average. \eagle\ shows that $j_{\rm stars}$ 
continues to rise out to much larger 
radii due to the effect of adding halo stars. \eagle\ predicts that this difference shrinks at larger radii, although still not 
converging to a fraction of $j_{\rm halo}$ as high as galaxies with no mergers in their lifetime. 
Early simulations of mergers predicted that dynamical friction could 
redistribute AM from the inner to the outer regions (e.g. \citealt{Barnes87,Navarro94,Heyl96}). From those simulations one 
would expect a net weak conservation of $j$. Our findings with \eagle\ show a significant disparity between the 
stellar and the halo $j$, but 
that is not as large as suggested by some of the idealised models \citep{Romanowsky12}.

\begin{figure}
\begin{center}
\includegraphics[trim=3mm 4mm 0mm 4mm, clip,width=0.5\textwidth]{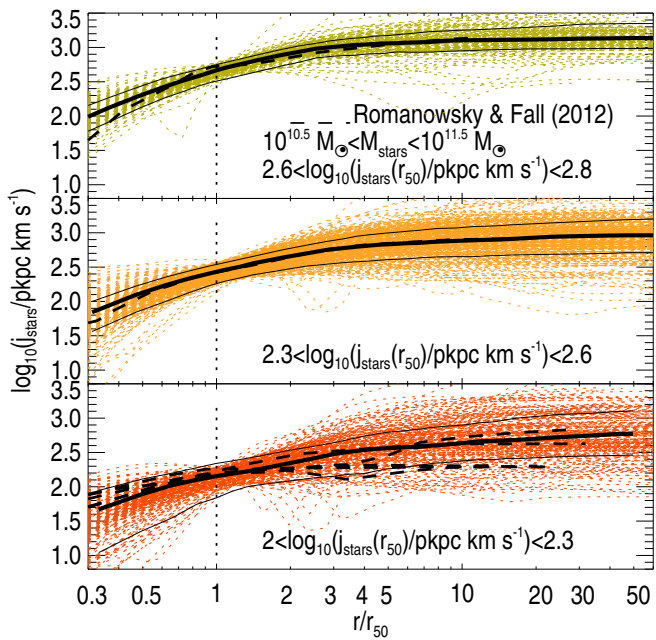}
\caption{Mean radial $j_{\rm stars}$ profiles for galaxies in \eagle\ at $z=0$ and with stellar masses in the range $10^{10.5}\,\rm M_{\odot}-10^{11.5}\,\rm M_{\odot}$ 
in $3$ bins of $j_{\rm stars}(r_{50})$, as labelled in each panel. In dotted lines we show all galaxies in that range, while the thick and thin solid lines show the median, and the 
$16^{\rm th}$ and $84^{\rm th}$ percentile ranges, respectively. We show observations of early-type galaxies from \citet{Romanowsky12} as dashed lines. Their sampled galaxies  
 have stellar masses in the range we selected here, and we show each galaxy in their corresponding bin of $j_{\rm stars}(r_{50})$. Here we only show the 
median measurement, but errorbars around those measurements can be as large as $\approx 0.5$~dex, particularly at $r\gtrsim 3r_{50}$.}
\label{obscomp}
\end{center}
\end{figure}
\begin{figure}
\begin{center}
\includegraphics[trim=3mm 4mm 0mm 4mm, clip,width=0.5\textwidth]{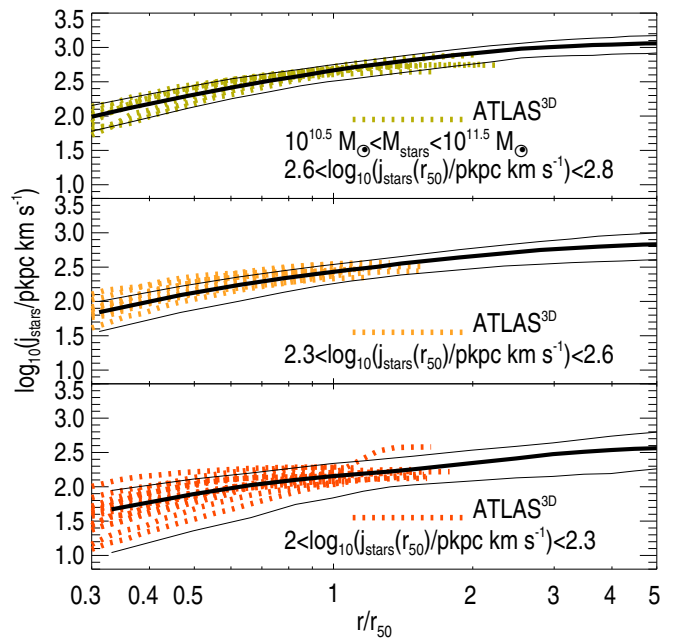}
\caption{Mean radial $j_{\rm stars}$ profile for the same \eagle\ galaxies of Fig.~\ref{obscomp}. For the simulated galaxies, we show the median and the 
$16^{\rm th}$ and $84^{\rm th}$ percentile ranges. Individual dotted lines show ATLAS$^{\rm 3D}$ galaxies 
that have stellar masses in the range $10^{10.5}\,\rm M_{\odot}-10^{11.5}\,\rm M_{\odot}$ and have $j_{\rm stars}(r_{50})$ in the ranges shown in each panel. 
The agreement between \eagle\ and the observations is excellent.}
\label{obscomp2}
\end{center}
\end{figure}

\subsubsection{Comparison with observations of early-type galaxies}

We compare \eagle\ galaxies of 
low $j_{\rm stars}$ with the observations of \citet{Romanowsky12} in Fig.~\ref{obscomp}. 
\citet{Romanowsky12} presented mean radial $j_{\rm stars}$ profiles for $7$ ellipticals and S0 galaxies in the stellar mass range 
of $10^{10.5}\,\rm M_{\odot} \lesssim M_{\rm stars}\lesssim 10^{11.5}\,\rm M_{\odot}$. We took these $7$ galaxies and separated them into $3$ 
bins of ${\rm log}_{10}(j_{\rm stars}(r_{50})/\rm pkpc\,km\,s^{-1})$, $2-2.3$, $2.4-2.6$ and $2.6-2.8$, which in \eagle\ would 
correspond to galaxies 
below, close to and above the median $j_{\rm stars}(r_{50})$ at that stellar mass, 
and compare them with \eagle\ in Fig.~\ref{obscomp}. In \eagle\ most galaxies of such stellar mass
are expected to be early-type (see Figs.~\ref{VisualJstars}~and~\ref{Nsersic}).  
We find that at 
low $j_{\rm stars}$ (bottom panel of Fig.~\ref{obscomp}) the scatter in the mean radial profiles becomes increasingly larger
compared to galaxies of higher $j_{\rm stars}$, and galaxies with flat mean $j_{\rm stars}$ profiles become more common. 
The diversity of mean radial $j_{\rm stars}$ profiles observed by 
\citet{Romanowsky12} is well captured by \eagle, even in the cases where $j_{\rm stars}$ ceases 
to increase at $r\gtrsim 3\,r_{50}$. 

With the aim of testing \eagle\ with a larger number of galaxies, we extracted mean radial $j_{\rm stars}$ profiles 
for every ATLAS$^{\rm 3D}$ galaxy \citep{Cappellari11}, following the procedure described in \citet{Lagos16b}. These profiles 
sample up to $\approx 2\,r_{\rm 50}$ in the best cases, but they inform us of the shape of the 
radial $j_{\rm stars}$ profile in the inner regions of galaxies. Fig.~\ref{obscomp2} shows the comparison 
between \eagle\ and ATLAS$^{\rm 3D}$ 
in the same stellar mass and $j_{\rm stars}(r_{50})$ ranges of Fig.~\ref{obscomp}. From top to bottom, each panel shows 
$8$, $10$ and $15$ ATLAS$^{\rm 3D}$ galaxies, respectively. The agreement with the observations is excellent. 
ATLAS$^{\rm 3D}$ galaxies show a larger scatter in the radial $j_{\rm stars}$ profiles with decreasing galaxy's spins, 
which is very similar to the trend seen in \eagle. This gives us confidence that the 
simulation not only reproduces the average $j$-mass relation, as shown by \citet{Lagos16b}, 
but also the radial profiles of galaxies, where observations exists.
The number of galaxies in the Universe in which this comparison can be done is still very sparse, but the advent of 
IFU instruments (e.g. SAMI, MaNGA, MUSE) is likely to change this.

\subsection{$j_{\rm stars}$ before and after galaxy mergers}\label{temporalevo}

\begin{figure*}
\begin{center}
\includegraphics[trim=0mm 24.5mm 0mm 1mm, clip,width=0.89\textwidth]{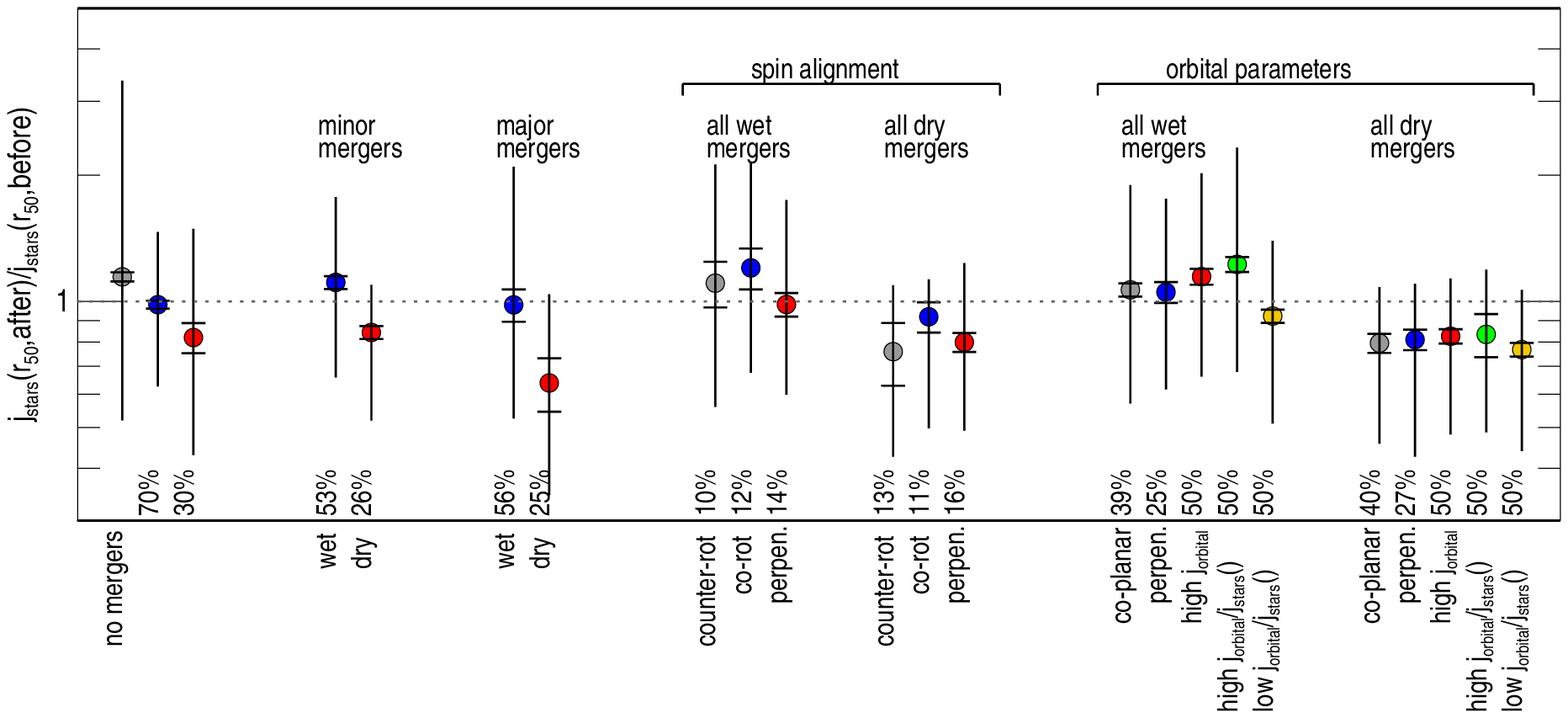}
\includegraphics[trim=0mm 0mm 0mm 1mm, clip,width=0.89\textwidth]{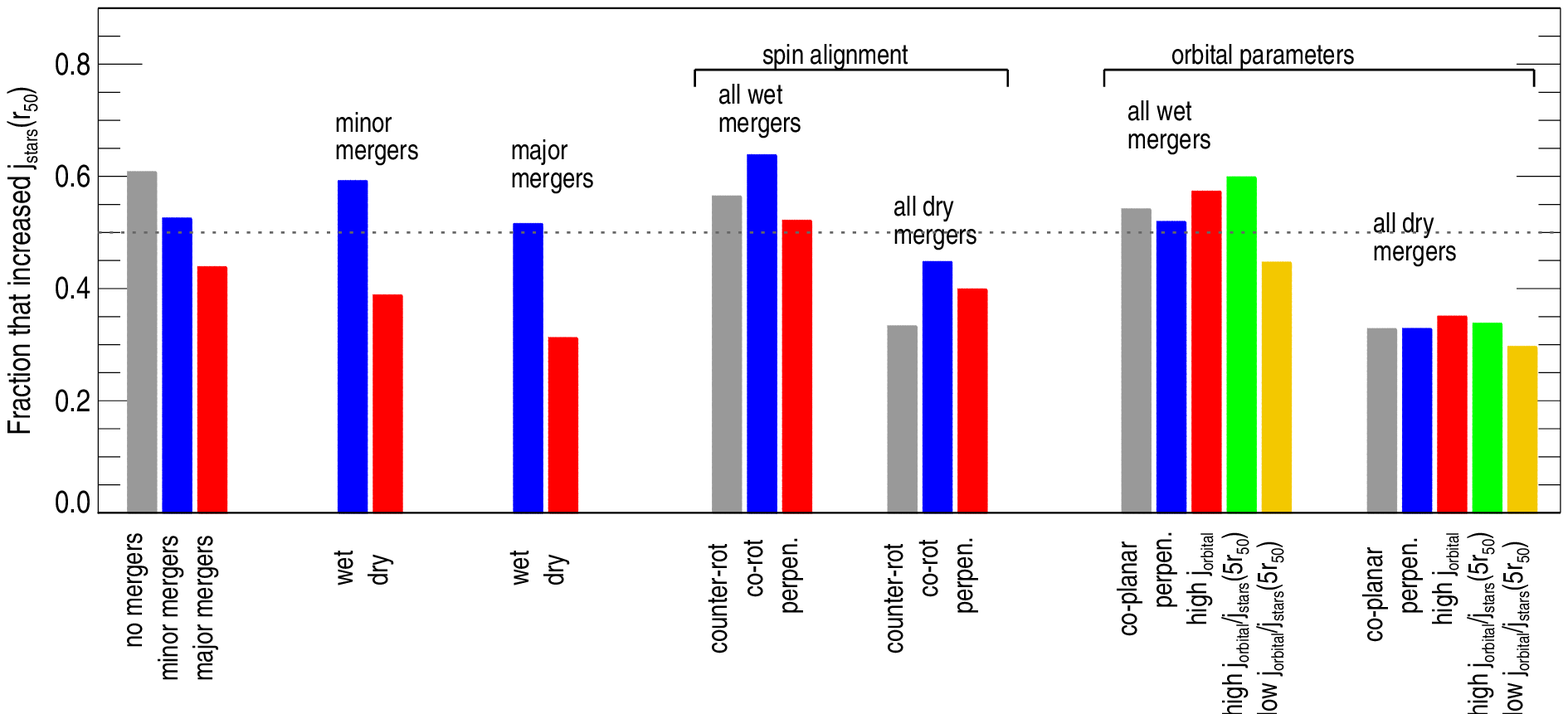}
\caption{{\it Top panel:} the ratio of $j_{\rm stars}(r_{\rm 50})$ in a galaxy between two consecutive snapshots. 
We separate galaxies into those that had no mergers, and those that went through a minor or a major merger (first 3 symbols); 
 galaxies that went through a minor or major merger, separated into wet and dry (subsequent 4 symbols); 
galaxies that went through wet and dry mergers separated into three samples of spin alignment (counter-rotating, co-rotating and perpendicular, as defined in
$\S$~\ref{EagleSec}; subsequent 6 symbols), and separated into $5$ orbital parameter samples (co-planar and perpendicular mergers,
and mergers with high orbital AM, and high/low orbital-to-central galaxy AM; subsequent 10 symbols), as labelled 
in the $x$-axis. The symbols show the medians, while the small and large errorbars show the statistical uncertainty on the median (from bootstrap resampling) 
and the $25^{\rm th}-75^{\rm th}$ percentile ranges, respectively.
The dotted line shows no change in $j_{\rm stars}(r_{\rm 50})$.
At the bottom of the panel we show the percentage of the mergers that are included in each subsample. 
{\it Bottom panel:} Fraction of galaxies that display an increase in their $j_{\rm stars}(r_{\rm 50})$ in the same cases shown for the same selections of the top panel.
For reference, the dotted line shows a fraction of $0.5$. 
We find that on average 
galaxy mergers have a negative effect on $j_{\rm stars}(r_{\rm 50})$, as a smaller fraction leads to an increase in $j_{\rm stars}(r_{\rm 50})$ 
compared to smooth accretion. However, the nature of the merger has a large effect on the outcome: wet, co-rotating mergers tend to increase 
$j_{\rm stars}(r_{\rm 50})$, while dry, counter-rotating mergers have the most negative effect on $j_{\rm stars}(r_{\rm 50})$.}
\label{HistogramsDeltaJ}
\end{center}
\end{figure*}

We analyse how much $j_{\rm stars}(r_{50})$ changes between two consecutive snapshots for galaxies with $M_{\rm stars}\ge 10^{9.5}\,\rm M_{\odot}$ 
and in the redshift range $0\le z\le 2.5$, separating galaxies into those that had and did not have a merger, and splitting mergers into different 
types: minor/major, wet/dry and with different spin alignments and orbital parameters. This is shown in the top 
panel of Fig.~\ref{HistogramsDeltaJ} (the equivalent for $j_{\rm stars}(5r_{50})$ is shown in Fig.~\ref{HistogramsDeltaJcMass}). 
The bottom panel of Fig.~\ref{HistogramsDeltaJ} shows the fraction of galaxies displaying 
an increase in $j_{\rm stars}(r_{50})$ for the same 
cases analysed in the top panel. 
The first $3$ data points compare the change in $j_{\rm stars}(r_{50})$ due to smooth accretion 
and a minor or major merger. In the latter case, star formation and gas accretion 
may continue, and thus, we cannot fully separate this effect from the merger.
Galaxies that did not suffer mergers on average increase their $j_{\rm stars}(r_{50})$ by $\approx 15$\% in between snapshots, and  
are likely to undergo an important increase in $j_{\rm stars}(r_{50})$ (i.e. $\approx 35$\% of the galaxies at least double their 
$j_{\rm stars}(r_{50})$ in a snapshot). On the contrary, galaxy mergers are more likely to not change or reduce 
their $j_{\rm stars}(r_{50})$, depending 
on whether they are minor or major mergers, respectively. 
From the bottom panel of Fig.~\ref{HistogramsDeltaJ}, one sees that smooth accretion increases 
$j_{\rm stars}(r_{50})$ $\approx 60$\% of the time, 
while minor and major mergers only do this in $\approx 54$\% and $\approx 43$\% of the cases, respectively.

Fig.~\ref{HistogramsDeltaJ} 
also splits mergers into several subsamples to pin down the circumstances in which $j_{\rm stars}$ change the most. 
We first take all of the minor and major mergers and split them into dry and wet (shown 
from the $4$th to the $7$th symbols and bars in Fig.~\ref{HistogramsDeltaJ}). We find that wet minor mergers 
produce a similar increase of $j_{\rm stars}(r_{50})$ to smooth accretion, with a smaller percentage of galaxies 
going through a major increase in $j_{\rm stars}(r_{50})$ ($\approx 20$\% of wet minor mergers produce 
an increase of a factor of $\gtrsim 2$). 
Dry minor mergers, on the other hand, display a strong preference for decreasing $j_{\rm stars}(r_{50})$. 
For major mergers the trends are similar but with a larger difference 
between dry and wet mergers. Dry major mergers reduce $j_{\rm stars}(r_{50})$ in $\approx 75$\% of the cases, 
which shows that this is one of the most catastrophic forms of mergers 
for the AM budget of galaxies. Note that in \eagle\ {\it the gas fraction of the merger is more important than the mass ratio}.
We calculate the Kolmogorov-Smirnov p-value between dry and wet mergers for the cases of minor and major mergers 
and find that there is negligible probability, $<10^{-15}$, that
they are drawn from the same population. 

So far we have stacked all of the galaxy mergers that take place in galaxies with $M_{\rm stars}\ge 10^{9.5}\,\rm M_{\odot}$
and in the redshift range $0\le z\le 2.5$. This may introduce significant biases due to the time 
interval between outputs of the 
simulation (different snapshots cover different timescales), and also due to galaxies having very different sizes at different cosmic epochs.
In Appendix~\ref{ExtraTests} we show that the bias introduced by studying mergers at different cosmic epochs and taking place in 
galaxies of different stellar masses is minimal, and that the difference seen here between minor/major, wet/dry mergers is of the same 
magnitude in subsamples of different redshifts and stellar masses.
From here on, we analyse galaxy mergers together, regardless the cosmic epoch and the stellar mass of the galaxy in which 
they occur, unless otherwise stated.

Given the importance of wet/dry mergers over minor/major mergers, 
we explore the effect of spin alignments and orbital parameters in the subsamples of 
dry and wet mergers in the right part of Fig.~\ref{HistogramsDeltaJ}. We first measure the effect of 
co-rotating ($\rm cos(\theta_{\rm spin})>0.7$), perpendicular ($-0.15<\rm cos(\theta_{\rm spin})<0.15$) and 
counter-rotating ($\rm cos(\theta_{\rm spin})<-0.7$) mergers (middle symbols and bars in Fig.~\ref{HistogramsDeltaJ}). 
We find that wet mergers between co-rotating galaxies lead to a larger and more frequent increase of $j_{\rm stars}(r_{50})$, while 
perpendicular wet mergers tend to produce little changes in $j_{\rm stars}(r_{50})$. 
$64$\% of the co-rotating wet minor mergers increase $j_{\rm stars}(r_{50})$, a frequency that is even higher than smooth accretion.
The effect of counter-rotating mergers 
is in between the co-rotating and perpendicular mergers. Perpendicular mergers are the most common configuration in \eagle\ 
(see Fig.~\ref{FrequencyMergers}) and that is why the bars for wet minor and major mergers are skewed towards the 
results of perpendicular rather than co-rotating mergers.
For dry mergers we find the same trend: co-rotating mergers tend to be less damaging than perpendicular or counter-rotating mergers 
for $j_{\rm stars}(r_{50})$.

In the rightmost part of Fig.~\ref{HistogramsDeltaJ} we analyse the effect of the orbital parameters. Particularly, we analyse 
co-planar and perpendicular mergers, the subsample with high $j_{\rm orbital}$ (i.e. higher than the median), and 
with high and low $j_{\rm orbital}/j_{\rm stars}(5r_{50})$ (above and below the median, respectively). 
$\S$~\ref{mergerpars} presents the definition of $j_{\rm orbital}$, and here we compare $j_{\rm orbital}$ to $j_{\rm stars}(5r_{50})$ 
of the primary galaxy prior to the merger. $j_{\rm stars}(5r_{50})$ is a good measurement of the galaxy's total 
$j_{\rm stars}$ (see Fig.~\ref{CumulativeJstars}). 
We do not find a strong effect of the orientation of the mergers on $j_{\rm stars}(5r_{50})$, as both the distributions of the 
co-planar and perpendicular mergers are statistically indistinguishable (the KS $p$-value is $0.56$).
When comparing mergers of high and low $j_{\rm orbital}$, however, we find a significant difference (with a KS $p$-value of $10^{-5}$) 
in which mergers with high $j_{\rm orbital}$ preferentially result in an increase in $j_{\rm stars}(5r_{50})$ 
of $\approx 15$\%. The largest systematic is found when we separate wet mergers by their $j_{\rm orbital}/j_{\rm stars}(5r_{50})$ 
(the $p$-value comparing the two subsamples of high/low $j_{\rm orbital}/j_{\rm stars}(5r_{50})$ is $10^{-24}$).
High values of $j_{\rm orbital}/j_{\rm stars}(5r_{50})$ efficiently spin-up the galaxy, increasing $j_{\rm stars}(r_{50})$ 
by $\approx 22$\%, on average, and in $60$\% of the cases. This suggests that galaxies spin-up because part of the 
orbital AM is transferred to the remnant galaxy. 
We study the subsample of wet, co-rotating ($\rm cos(\theta_{\rm spin})>0.7$) and high $j_{\rm orbital}/j_{\rm stars}(5r_{50})$ mergers, and 
find that they increase $j_{\rm stars}(r_{50})$ in $\approx 70$\% of the cases, by $\approx 44$\% on average, and thus this
form of merger is the most efficient at spinning-up galaxies.
In the case of dry mergers we do not find a strong dependence on any of the orbital parameters studied here.

When studying $j_{\rm stars}(5r_{50})$ (Fig.~\ref{HistogramsDeltaJ5}) we find very similar results.
The only major difference is that dry mergers show a stronger dependence on the orbital parameters, 
with high $j_{\rm orbital}/j_{\rm stars}(5r_{50})$ and 
co-planar mergers leading to a higher fraction of galaxies displaying and increase in $j_{\rm stars}(5r_{50})$. Thus, we conclude that the AM
in the inner parts of galaxies during dry mergers is not greatly affected by the orbital parameters of the mergers, but when focusing 
on the total $j_{\rm stars}$, we see that perpendicular and low $j_{\rm orbital}/j_{\rm stars}(5r_{50})$ mergers, are the most catastrophic.  

We conclude that in \eagle\, wet, co-rotating mergers can spin-up galaxies very efficiently, 
and even more if they have a high $j_{\rm orbital}/j_{\rm stars}(5r_{50})$. 
On the contrary, dry, counter-rotating mergers are the most effective at spinning down galaxies.
The environment in which these mergers take place may have a significant impact. We find that wet mergers 
generally happen in halos of higher spins compared to the 
median of all halos. This could be interpreted as accretion spinning up halos, 
as well as making the galaxies gas-rich and resulting in a high spin merger remnant. 
The consequences of such correlation are very interesting but beyond the scope of this paper, 
so we defer it to future investigation.

\subsection{Rearrangement of $j_{\rm stars}$ during galaxy mergers}\label{rearr-Sec}

\begin{figure}
\begin{center}
\includegraphics[trim=0mm 1mm 0mm 1mm, clip,width=0.5\textwidth]{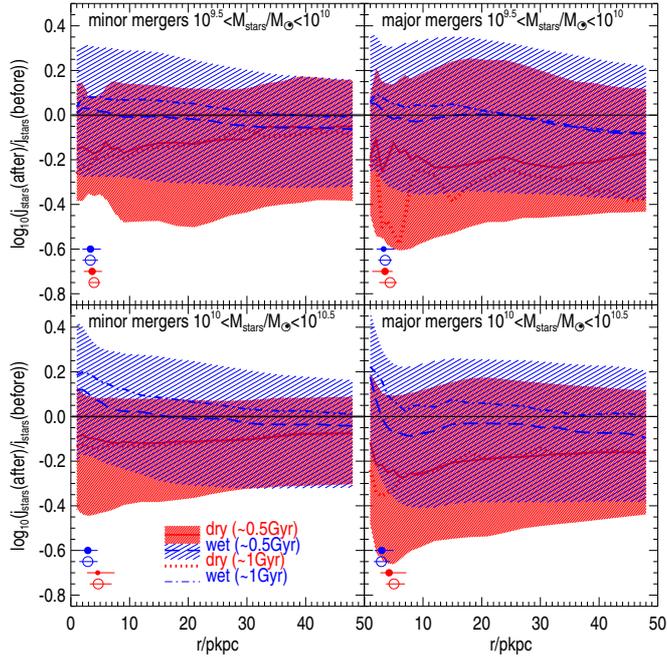}
\caption{The ratio between the mean radial $j_{\rm stars}$ profiles 
after and before the galaxy merger, measured in an aperture $r$, as a function of $r$.
We measure $j_{\rm stars}$ post-merger in the snapshot right after the merger, and two snapshots after, 
which correspond approximately to $0.5$ and $1$~Gyr after the merger, respectively.
Minor and major mergers are shown in the left and right panels, respectively, in two bins of the neutral gas fraction of the merger, as labelled.
The top panels show galaxies with $10^{9.5}\,\rm M_{\odot}<M_{\rm stars}<10^{10}\,\rm M_{\odot}$, while the bottom panels show 
galaxies with $10^{10}\,\rm M_{\odot}<M_{\rm stars}<10^{10.5}\,\rm M_{\odot}$.
Lines and the shaded regions show the median and the $25^{\rm th}$-$75^{\rm th}$ percentile ranges.
The latter are calculated using the snapshots right after the merger.
The filled and empty circles with the error bar at the bottom of each panel show the median $r_{\rm 50}$ before and after the merger, respectively, for each sample and the 
$25^{\rm th}$-$75^{\rm th}$ percentile range, respectively. 
Horizontal lines mark no change in $j_{\rm stars}(r)$, and so values above (below)
correspond to $j_{\rm stars}$ increasing (decreasing).
The figure reveals that wet mergers tend to increase $j_{\rm stars}$ in the 
inner regions of galaxies, while decreasing it in the outer regions.}
\label{DeltaJDeltaM_minormajorc}
\end{center}
\end{figure}


\begin{figure}
\begin{center}
\includegraphics[trim=0mm 4.5mm 0mm 1mm, clip,width=0.4\textwidth]{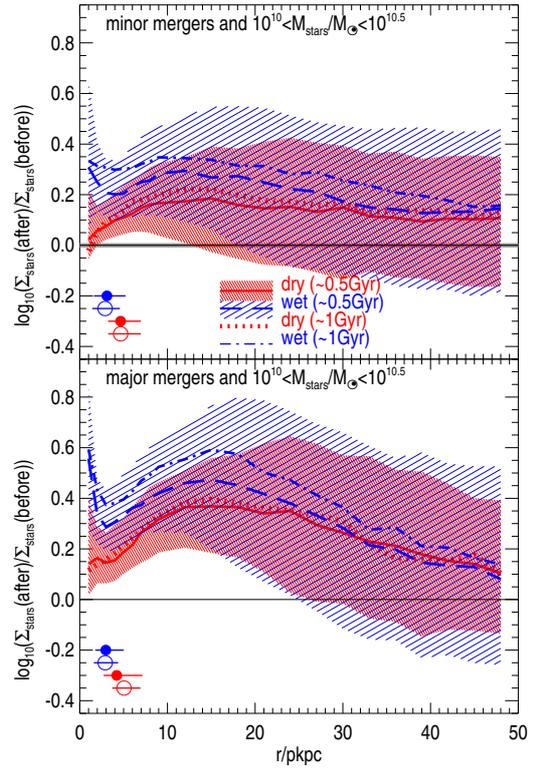}
\caption{Stellar mass surface density profile 
before and after the mergers for the galaxies in the bottom panel of Fig.~\ref{DeltaJDeltaM_minormajorc}. 
This figure shows that gas-rich mergers tend to build the central stellar mass concentration (i.e. bulge), while dry mergers increase the stellar mass density towards the outskirts of galaxies. 
the latter case usually drives an increase in $r_{\rm 50}$, while the former does not change $r_{\rm 50}$ significantly.}
\label{SurfaceStellarMass}
\end{center}
\end{figure}

In Fig.~\ref{DeltaJDeltaM_minormajorc} we study the mean radial $j_{\rm stars}$ profile of the primary galaxy before and after the merger 
in two bins of stellar mass and for minor/major mergers that are wet/dry. {\it Before} the merger here means the last snapshot 
in which the galaxy participating in the merger was individually identified, and for `{\it after}' the merger we look at the two 
consecutive snapshots in which the galaxies has been identified as one (i.e. already merged in the merger tree). Given the time period 
in between snapshots, the two consecutive snapshots roughly correspond to $\approx 0.5$~Gyr and $\approx 1$~Gyr, respectively, after the merger.
We study two snapshots after the merger because visual inspection of mergers in \eagle\ reveals that in some cases the 
{merger tree algorithm} 
considers a galaxy pair as already merged even though the process is still ongoing. 
Another motivation to study two consecutive snapshots after the 
merger is to test the effect of relaxation if any is present. 

In the low stellar mass bin 
of Fig.~\ref{DeltaJDeltaM_minormajorc}, we show that both dry minor and 
major mergers have the effect of reducing $j_{\rm stars}$ across the entire radii range considered. 
Studying $j_{\rm stars}$ at $\approx 0.5$ or $\approx 1$~Gyr after the merger makes 
little difference in this case. Major dry mergers tend to reduce $j_{\rm stars}$ by $\approx 0.2$~dex on average in both 
low and high stellar mass bins, while
minor dry mergers drive a more modest reduction of $\approx 0.1$~dex, on average.
In the case of wet mergers we see a differential effect on the $j_{\rm stars}$ 
profiles: inner regions of galaxies, $r\lesssim 5$~kpc (typically $\approx 2 r_{50}$; see 
filled and open circles in Fig.~\ref{DeltaJDeltaM_minormajorc}), tend to increase 
 $j_{\rm stars}$, while at larger radii, $j_{\rm stars}$ tends to decrease if one looks at the merger 
remnant $\approx 0.5$~Gyr after the merger, or very modestly increase 
if studied $\approx 1$~Gyr after. The latter could be due to a combination of relaxation and 
continuing gas accretion and star formation. Separating the latter 
is not obvious in a simulation like \eagle\ where all the physical processes are interplaying at any given time.

\subsubsection{The physical origin of the $j_{\rm stars}$ increase in wet mergers}\label{WetMersDetails}

To further understand the differential effect wet mergers have on the 
mean radial $j_{\rm stars}$ profile, we study in Fig.~\ref{SurfaceStellarMass} 
the change in the stellar surface density of the primary galaxy before and after the merger. 
For clarity, we only plot the mass bin $10^{10}\,\rm M_{\odot}<M_{\rm stars}<10^{10.5}\,\rm M_{\odot}$ 
as the lower mass bin gives very similar results.
Fig.~\ref{SurfaceStellarMass} shows that wet major and minor mergers drive a significant increase in the central stellar surface density by 
a factor 
$\gtrsim 0.2$~dex, on average. At intermediate radii, 
$5\rm \,pkpc\lesssim r\lesssim 30\rm \,pkpc$ there is also an increase, but of a less significant magnitude.
If the central stellar mass (i.e. bulge) is increasing, 
and the rotational velocity increases as $v \approx \sqrt{GM/r}$, 
$j_{\rm stars}$ is also expected to increase.
This effect has been seen before in non-cosmological simulations of gas-rich mergers 
(\citealt{Springel00}; \citealt{Cox06}; \citealt{Robertson06}; \citealt{Johansson09}; \citealt{Peirani10}; \citealt{Moreno15}).

\begin{figure}
\begin{center}
\includegraphics[trim=2mm 4mm 0mm 5mm, clip,width=0.499\textwidth]{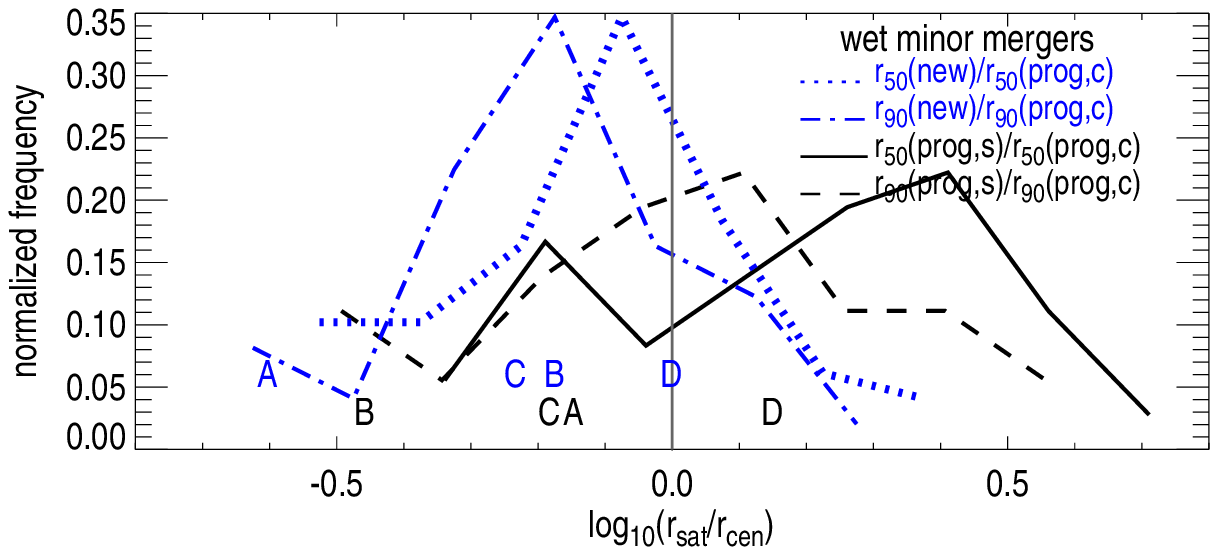}
\includegraphics[trim=0mm 0mm 0mm 0mm, clip,width=0.479\textwidth]{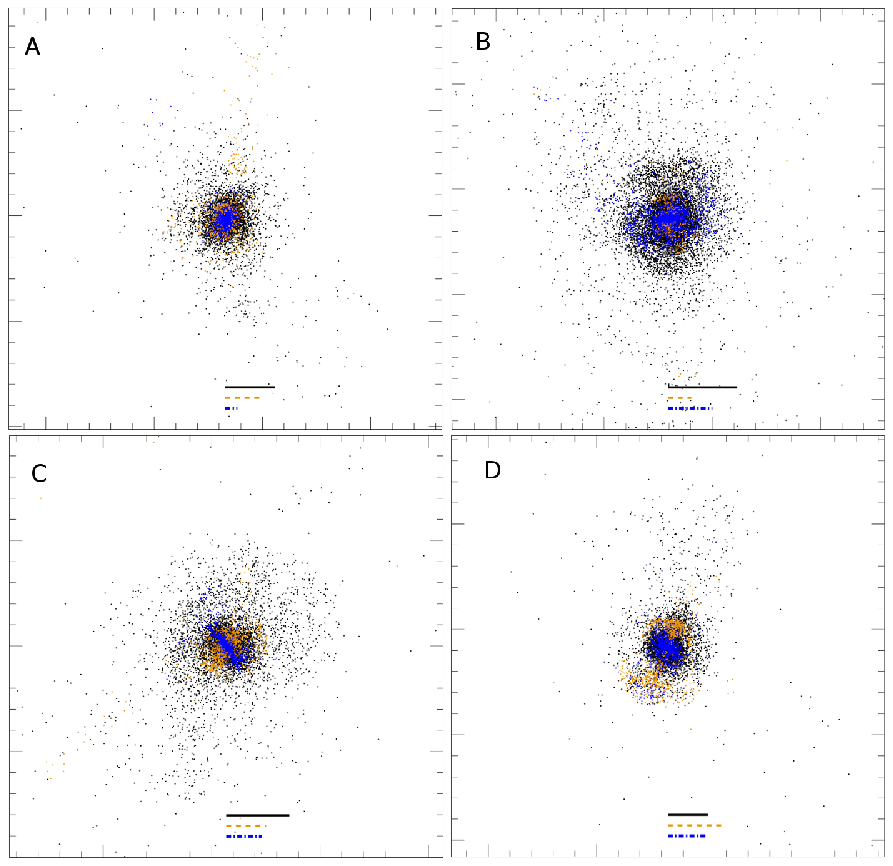}
\caption{{\it Top panel:} The ratio between the $50$\% and $90$\% stellar mass radii of the progenitor 
satellite (labelled as `prog,s') and the newly formed stars vs. the progenitor central stellar (labelled as `prog,c') 
components, as labelled,
in all wet minor mergers in the redshift range $\approx 0.2-0.8$ and that took place in primary galaxies with
$M_{\rm stars}\ge 10^{9.5}\,\rm M_{\odot}$ in \eagle.
This figure shows that newly formed stars reside in the centre of the galaxy, and are more concentrated
than the stars that were in primary galaxy before the merger. 
{\it Bottom panel:} stellar-particle distribution in $4$ examples of wet major mergers that span the range of size ratios shown in the top panel.
The images are $x$-$y$ projections of $200$~ckpc on a side. Black, yellow and blue points show progenitor stars that belonged to the primary galaxy,
progenitor stars that belonged to the secondary galaxy and stars that formed during the merger, respectively. The segments of the same colours at the bottom
show $r_{90}$ of the three components.}
\label{MergersSizes2}
\end{center}
\end{figure}

\begin{figure}
\begin{center}
\includegraphics[trim=2mm 3mm 0mm 5mm, clip,width=0.45\textwidth]{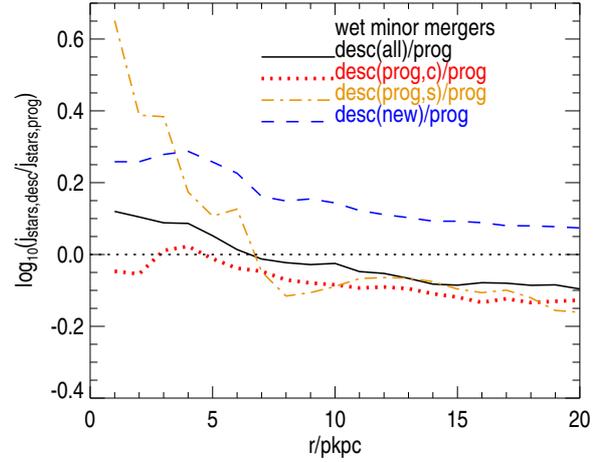}
\caption{The ratio between the mean radial $j_{\rm stars}$ profiles
after and before the galaxy merger (solid line), measured in an aperture $r$, as a function of $r$, 
for all wet minor mergers in the redshift range $\approx 0.2-0.8$ and that took place in primary galaxies with
$M_{\rm stars}\ge 10^{9.5}\,\rm M_{\odot}$ in \eagle.
We also show the ratio between the mean radial $j_{\rm stars}$ profiles of the descendant and the 
progenitor but when we measure the former only with the 
progenitor central (dotted line), progenitor satellite (dot-dashed line) and newly formed stars (dashed line). 
Here lines show the medians. This figure 
shows that the increase in $j_{\rm stars}$ in a wet merger is due to the  contribution of newly formed stars.}
\label{JMergersSizes2}
\end{center}
\end{figure}


One remaining question is whether the build-up of the bulge is driven by a preferential deposition 
of the stars of the satellite galaxy in the centre, 
by dynamical friction moving stars of the primary galaxy to the centre, or the preferential formation of new stars in the centre.
To answer this question we identified in the merger remnant the stars that belonged to the secondary 
(i.e. progenitor satellite stars) 
and primary (i.e. progenitor central stars)
galaxy before the merger, and those that formed
during the merger (i.e. new stars), and calculate their $50$\% and $90$\% stellar mass radii.
We do this for all mergers that took place in the redshift range $\approx 0.2-0.8$, 
which is of particular interest, as it is the time when the universe 
goes from being dominated by wet to dry mergers in \eagle\ (see Fig.~\ref{FrequencyMergers}). 
Fig.~\ref{MergersSizes2} shows the ratio of $r_{50}$ and $r_{90}$ between the progenitor satellite stars and the progenitor 
central stars, and between the 
new stars and the progenitor central stars in the case of wet minor mergers. For the new stars, we find that in $\approx 73$\% of 
cases they end up more concentrated 
and with $r_{50}$ and $r_{90}$ typically $\approx 1.3$ times smaller than the progenitor central stars. 
For the progenitor satellite stars, we find that in $\approx 70$\% of the cases they end up more extended 
and with $r_{50}$ and $r_{90}$ values that are $\approx 1.8$ and $\approx 1.3$ times larger than those of the progenitor central stars.
The bottom panel of Fig.~\ref{MergersSizes2} shows $4$ examples of wet minor merger remnants and how the stars 
from the $3$ components above are spatially distributed. 
We generally find that when $r_{\rm 50}$ of the progenitor satellite stars is larger than that of the progenitor
 central stars 
there is an associated extended stellar structure in the form of streams or shells (e.g. galaxy `D' in Fig.~\ref{MergersSizes2}). 
If we focus on the central $2$~pkpc, we find that the bulge mass is dominated by
the progenitor central stars ($\approx 70$\% on average), but with a large contribution from the newly formed stars ($\approx 30$\% on average).
Although there is a significant contribution of newly formed stars, we find that the mass-weighted age of the bulge 
by $z=0$ is $\gtrsim 9.5$~Gyr old, on average, 
due to the stars contributed by the primary and secondary galaxies that end up in the central $2$~pkpc being extremely old.

We also studied the contribution of these stars to the mean $j_{\rm stars}$ radial profile of the merger remnant in the inner $20$~pkpc 
in Fig.~\ref{JMergersSizes2}. 
We find that the increase in $j_{\rm stars}$ in the inner regions of galaxies as a result of the wet merger and 
reported in Fig.~\ref{DeltaJDeltaM_minormajorc} is due to the newly formed stars. 
Although the progenitor satellite stars also have a high $j_{\rm stars}$ compared to the progenitor central stars, 
their contribution to the stellar mass is very small. In fact, in the inner $2$~pkpc, newly formed stars are responsible for 
$33$\% of the $j_{\rm stars}$ of the descendant, while progenitor central and satellite stars contribute $58$\% and $9$\%, respectively, on average.
At larger radii, $j_{\rm stars}$ of the descendant is dominated by the stars of the progenitor central galaxy.

The main difference between wet minor and major mergers, in that in the latter (not shown here) 
the stars belonging to the progenitor secondary galaxy end up more concentrated than the 
progenitor central stars (typically $\approx 1.5$ times more concentrated, on average). 

We conclude that the increase of $j_{\rm stars}$ in the inner regions of galaxies 
as a result of a wet merger is caused primarily 
by the flows of gas towards
the centre that subsequently form stars.
{These new stars contribute to the formation of the bulge, and are typically characterised by 
higher $j$ at fixed radius, thus, producing steeper velocity profiles}.

\subsubsection{The effect of spin and orbital alignments}

\begin{figure}
\begin{center}
\includegraphics[trim=0mm 2mm 0mm 1mm, clip,width=0.5\textwidth]{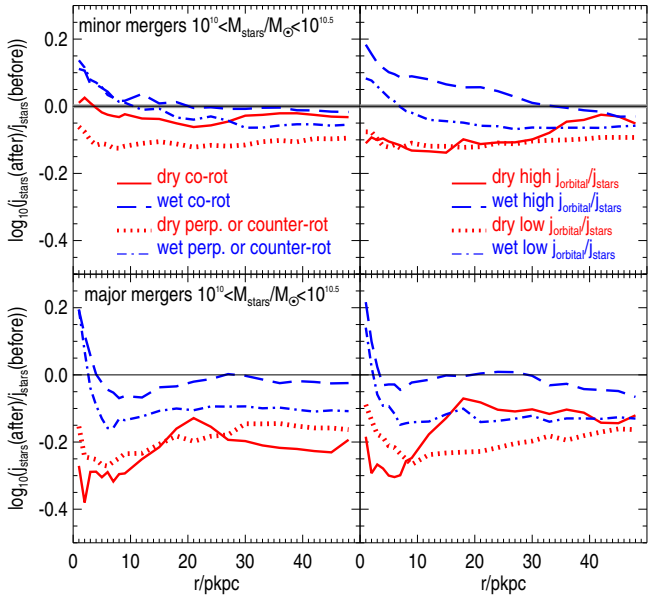}
\caption{The ratio between the mean radial $j_{\rm stars}$ 
profiles after and before the galaxy merger, measured in an aperture $r$, as a function of $r$ 
for galaxies with stellar masses in the range $10^{10}\,\rm M_{\odot}<M_{\rm stars}<10^{10.5}\,\rm M_{\odot}$ at $z<2$. Mergers are split into minor (top panels) and 
major (bottom panels). In addition, every panel shows wet and dry mergers as red and blue lines split into 
co-rotating vs. perpendicular or counter-rotating mergers (left panels), high vs. low $j_{\rm orbital}/j_{\rm stars}$ (right panels), as labelled. 
Here we study consecutive snapshots, which in practice means that the profile {\it after} 
the merger is measured at $\approx 0.3-0.5$~Gyr after the merger. Lines correspond to the medians of the distributions. For 
clarity we do not show here the percentile ranges, but they are of a similar magnitude to those shown in Fig.~\ref{DeltaJDeltaM_minormajorc}.} 
\label{ProfilesROrbitalParameters}
\end{center}
\end{figure}

In Fig.~\ref{ProfilesROrbitalParameters} we show the mean radial $j_{\rm stars}$ 
profiles of galaxies before and after the merger. After the merger 
corresponds to 
the first snapshot in which the two merging galaxies are identified as one single remnant (typically $0.5$~Gyr after the merger). 
In the left panels of Fig.~\ref{ProfilesROrbitalParameters} we separate dry and wet minor (top) and major (bottom) mergers 
that took place in galaxies with $10^{10}\,\rm M_{\odot}<M_{\rm stars}<10^{10.5}\,\rm M_{\odot}$ into 
the subsamples of co-rotating ($\rm cos(\theta_{\rm spin})>0.7$; see Eq.~\ref{Eqspin} for a definition of $\theta_{\rm spin}$), 
and perpendicular or counter-rotating galaxies ($\rm cos(\theta_{\rm spin})<0.15$).
 
Wet minor mergers of galaxies that are co-rotating spin up the central region, due to the build-up of the bulge, 
and have very little effect on the outskirts of the galaxy (i.e. $j_{\rm stars}(\rm after)\sim j_{\rm stars}(\rm before)$). 
In the case of perpendicular or counter-rotating galaxies, there is a significant spin down at 
$r\gtrsim 10$~pkpc of $\approx 40$\%, on average, in the case of major mergers, and a more modest one of $\approx 12$\% 
for minor mergers.
A very significant difference is seen in dry minor mergers between co-rotating or perpendicular/counter-rotating galaxies. 
We find that very little happens to $j_{\rm stars}(r)$ if the dry minor merger is between co-rotating galaxies, while in the 
case of perpendicular/counter-rotating mergers, there is a significant stripping of $j_{\rm stars}(r)$ of $\approx 30$\%, on average, 
through the entire radii range studied here. Note that in the case of dry major mergers, there is always a significant stripping of 
$j_{\rm stars}$ regardless of the spin and orbital parameters.

In the right panels of Fig.~\ref{ProfilesROrbitalParameters} we show the ratio between the mean radial $j_{\rm stars}$ profiles 
before and after the merger 
as a function of $r$ for subsamples of dry/wet minor/major mergers, split into two bins of 
$j_{\rm orbital}/j_{\rm stars}(5r_{50})$. These two bins 
are above (high $j_{\rm orbital}/j_{\rm stars}(5r_{50})$) and below (low $j_{\rm orbital}/j_{\rm stars}(5r_{50})$) 
the median value of $j_{\rm orbital}/j_{\rm stars}(5r_{50})$. 
Here $j_{\rm stars}(5r_{50})$ corresponds to the value of the primary galaxy prior to
 the merger. In $\S$~\ref{temporalevo} we show that this was the most important orbital parameter 
determining whether a galaxy suffered a spin up or down as a result of the merger. 
In the case of high $j_{\rm orbital}/j_{\rm stars}(5r_{50})$
 we find that wet minor mergers result in a spin up that is significant out to $r\approx 30$~pkpc, 
increasing $j_{\rm stars}(r)$ by $\approx 60$\% at $r\lesssim 5$~pkpc and $\approx 25$\% 
at $5\rm \,pkpc\lesssim r\lesssim 15$~pkpc. Such a merger in \eagle\ is the most effective at spinning up galaxies.
These galaxies can end up in the upper envelope of the $j_{\rm stars}-M_{\rm stars}$ relation.
For wet major mergers, we find a significant increase in the very inner regions ($r\lesssim 3$~pkpc), and very little 
change at larger radii.
Dry mergers show very little difference between high and low $j_{\rm orbital}/j_{\rm stars}(5r_{50})$, on average.
\citet{DiMatteo09} showed that in the case of very high $j_{\rm orbital}$ the remnant can end up with a large 
$j_{\rm stars}$ even in the case of dry mergers. \eagle\ reveals that this type of event is very rare, and most of the 
time the galaxies spin down as a result of a dry merger. 

In the case of low $j_{\rm orbital}/j_{\rm stars}(5r_{50})$, wet mergers 
show modest to large losses of $j_{\rm stars}(<r)$.
This large difference between the high/low $j_{\rm orbital}/j_{\rm stars}(5r_{50})$ subsamples arise 
from the efficient transfer of $j_{\rm orbital}$ onto the remnant galaxy, which can significantly spin up 
a galaxy when $j_{\rm orbital}$ is large.

\section{Discussion and conclusions}\label{conclusions}

The classic interpretation of the positions of spiral and elliptical galaxies in the 
$j_{\rm stars}-M_{\rm stars}$ plane by e.g. \citet{Fall83} and \citet{Romanowsky12} says 
that spiral galaxies are the result of weak conservation of specific AM of the gas falling in 
 and forming stars, while elliptical galaxies loose 
$\gtrsim 50-90$\% of their $j$ during their formation process. 
The preferred invoked mechanism responsible for such loss is galaxy mergers. 

While we find mergers to preferentially spin galaxies down, 
their influence can be quite varied, and in 
many cases they spin galaxies up significantly, positioning them in the upper envelope of the $j_{\rm stars}$-stellar mass relation. 
The latter is the case of wet mergers between co-rotating galaxies and with high $j_{\rm orbital}$ relative to the $j_{\rm stars}$ of the 
galaxies prior to the merger. 
When studying the correlation between the positions of galaxies in the $j_{\rm stars}$-stellar mass plane and their merger history, 
we find the wet merger rate increases with decreasing stellar mass and increasing $j_{\rm stars}$, 
while the dry merger rate increases with increasing stellar mass and decreasing $j_{\rm stars}$. 
In fact, \eagle\ shows that for the $j_{\rm stars}$ value of the merger remnant galaxy, the most important parameter is the gas fraction of the 
merger, rather than the mass 
ratio or the spin/orbital parameters. The latter play a secondary, nonetheless relatively important, role.
Dry mergers are the most effective way of spinning down galaxies, though the subsample of minor, co-rotating mergers 
are relatively harmless. Counter-rotating dry mergers are the most efficient at spinning galaxies down.
Our definition of wet and dry is very gas-rich and gas-poor. 
Thus, dry 
mergers may be slightly different than the purely 
collisionless experiments widely discussed in the literature (e.g. 
\citealt{Boylan-Kolchin05}; \citealt{Naab06b}; \citealt{Taranu13}; \citealt{Naab14}). 

Classical results of dry mergers by early works (e.g. \citealt{Barnes87} and \citealt{Navarro97}), show that 
dynamical friction redistributes $j_{\rm stars}$ in a way such that most of it ends up at very large radii, but if integrating over a large enough 
baseline, one finds $j_{\rm stars}$ converging to $j_{\rm halo}$. These results were refuted by the observations of elliptical galaxies 
compiled by \citet{Romanowsky12}; these authors showed in a sample of $7$ early-type galaxies that some of them converged 
in their $j_{\rm stars}$ 
to values that would indicate a large deficiency compared to an average $j_{\rm halo}$. 
Using \eagle\ we found that dry merger remnants, 
those with the highest S\'ersic indices, have most of their $j_{\rm stars}$ 
budget at $r\gtrsim 5\times r_{50}$, in agreement with the early works discussed above, 
but that the variety of the radial $j_{\rm stars}$ profiles of galaxies, 
particularly at low $j_{\rm stars}(r_{50})$, can easily explain the rotation curves presented 
in \citet{Romanowsky12}. We compared the \eagle\ $j_{\rm stars}$ profiles with ATLAS$^{\rm 3D}$ 
galaxies and found excellent agreement. 
The main difference between what we find with \eagle\ and the early papers above, 
is that the total $j_{\rm stars}$ in the case of dry merger remnants 
converges to $\approx 20$\% of the halo $j$, on average, while galaxies that never had a merger 
of the same stellar mass, 
typically have a total $j_{\rm stars}$ that is $\approx 40$\% their $j_{\rm halo}$. Thus, a relatively modest 
but significant difference is found between these two samples.
 
The case of wet mergers in \eagle\ is very interesting from the perspective of 
$j_{\rm stars}$ and the morphology of galaxies. We find that 
in most of these mergers, the inner regions of galaxies undergo a spin up as a result of 
stars being formed in the central $\approx 2-5$~pkpc with high circular velocities.  
These newly formed stars are the result of gas inflows triggered by the merger, and 
drive the build-up of the bulge.
These new stars display a significantly more concentrated distribution compared to the stars that were present in the 
primary or the secondary galaxy before the merger. Stars that belonged to the secondary galaxy 
end up preferentially more concentrated than the stars of the primary galaxy in the case of 
major mergers, and significantly more extended in the case of minor mergers. 
These extended structures are in the form of streams and/or shells.

Key observational tests to support our findings for the effect of mergers on the $j_{\rm stars}$ 
of elliptical galaxies would be to increase the sample of elliptical galaxies with good kinematic information out to $10\,r_{50}$. 
Our predictions 
are: (i) the mean radial $j_{\rm stars}$ profiles of ellipticals are 
typically shallower than spiral galaxies, and that (ii) these profiles 
 continue to rise well beyond $10\,r_{50}$. 
A cautionary note: many of these stars that are beyond $10\,r_{50}$ would not necessarily 
be considered part of the galaxy, but instead 
they may belong to the stellar halo. In terms of the mean radial $j_{\rm stars}$ profile, 
however, we do not see obvious features that would indicate 
distinct stellar components.

A plausible strategy to test the raising $j_{\rm stars}$ profiles of ellipticals would be to use IFU surveys, 
such as SAMI and MaNGA, 
to define a suitable sample of galaxies, selected from the $j_{\rm stars}$-stellar mass plane, with $j_{\rm stars}$ 
here measured within some relatively small aperture (e.g. SAMI used one effective radius to measure $j_{\rm stars}$ within; \citealt{Cortese16}), 
and follow up to measure $j_{\rm stars}$ out to radii $>10\,r_{50}$. 
The latter can be achieved by studying the kinematics of planetary nebulae and/or globular clusters 
(e.g. \citealt{Coccato09}; \citealt{Romanowsky09}; \citealt{McNeil10}; \citealt{Foster11}). 
In addition, the lack of information on the $3$D stellar densities and velocities makes it necessary to 
develop fitting tools that enable the reconstruction of $3$D galaxies by imposing Newtonian constraints on IFU data. 
Observations and modelling tools like the ones described here 
would provide stringent constraints to the simulation and the galaxy formation physics included in it.

\section*{Acknowledgements}

We thank Luca Cortese and Matthieu Schaller for useful discussions and comments on the manuscript.
CL is funded by a Discovery Early Career Researcher Award (DE150100618).
CL also thanks the MERAC Foundation for a Postdoctoral Research Award.
This work was supported by a Research Collaboration Award at the University of Western Australia.
This work used the DiRAC Data Centric system at Durham University, operated by the Institute for Computational Cosmology on behalf of the STFC DiRAC HPC Facility ({\tt www.dirac.ac.uk}). This equipment was funded by BIS National E-infrastructure capital grant ST/K00042X/1, STFC capital grant ST/H008519/1, and STFC DiRAC Operations grant ST/K003267/1 and Durham University. DiRAC is part of the National E-Infrastructure.
Support was also received via the Interuniversity Attraction Poles Programme initiated
by the Belgian Science Policy Office ([AP P7/08 CHARM]), the
National Science Foundation under Grant No. NSF PHY11-25915,
and the UK Science and Technology Facilities Council (grant numbers ST/F001166/1 and ST/I000976/1) via rolling and
consolidating grants awarded to the ICC.
We acknowledge the Virgo Consortium for making
their simulation data available. The \eagle\ simulations were performed using the DiRAC-2 facility at
Durham, managed by the ICC, and the PRACE facility Curie based in France at TGCC, CEA, Bruyeres-le-Chatel.
This research was supported in part by the National Science Foundation under Grant No. NSF PHY11-25915.
Parts of this research were conducted by the Australian Research Council Centre of Excellence for All-sky Astrophysics (CAASTRO), through project number CE110001020, and 
supported by the Australian Research Council Discovery Project 160102235.

\bibliographystyle{mn2e_trunc8}
\bibliography{Mergers}

\label{lastpage}
\appendix
\section[]{Convergence test}\label{ConvTests}

\begin{table*}
\begin{center}
\caption{\eagle\ simulations used in this Appendix. The columns list:
    (1) the name of the simulation, (2) comoving box size, (3) number
    of particles, (4) initial particle masses of gas and (5) dark
    matter, (6) comoving gravitational
    softening length, and (7) maximum physical comoving Plummer-equivalent
    gravitational softening length. Units are indicated below the name of
    each column. \eagle\
    adopts (6) as the softening length at $z\ge 2.8$, and (7) at $z<2.8$.}\label{TableSimus2}
\begin{tabular}{l c c c c c l}
\\[3pt]
\hline
(1) & (2) & (3) & (4) & (5) & (6) & (7) \\
\hline
Name & $L$ & \# particles & gas particle mass & DM particle mass & Softening length & max. gravitational softening \\
Units & $[\rm cMpc]$   &                &  $[\rm M_{\odot}]$ &  $[\rm M_{\odot}]$ & $[\rm ckpc]$ & $[\rm pkpc]$\\
\hline
Ref-L025N0376 & $25$ &   ~$2\times 376^3$  &$1.81\times 10^6$   &   ~~$9.7\times 10^6$ & $2.66$ & ~~$0.7$  \\
Ref-L025N0752 & $25$ &   ~$2\times 752^3$  &$2.26\times 10^5$   &   $1.21\times 10^6$  & $1.33$ &  $0.35$ \\
\hline
\end{tabular}
\end{center}
\end{table*}

\begin{figure}
\begin{center}
\includegraphics[trim=0mm 4mm 0mm 4mm, clip,width=0.43\textwidth]{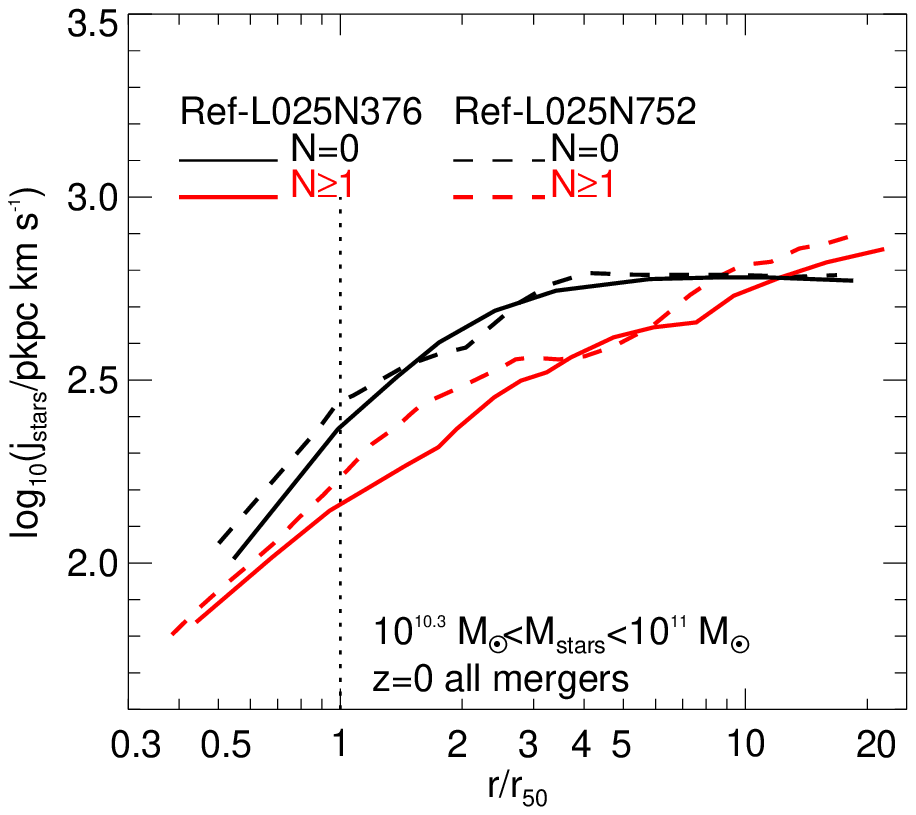}
\caption{$j_{\rm stars}(r)$ as a function of $r$ for galaxies 
with stellar masses in the range 
$10^{10.3}\,\rm M_{\odot}<M_{\rm stars}<10^{11}\,\rm M_{\odot}$ at $z=0$ 
for the Ref-L025N0376 (solid lines) and Ref-L025N0752 (dashed lines) simulations.
Lines show the median $j_{\rm stars}(r)$. The $16^{\rm th}$ to $84^{\rm th}$ percentile ranges are similar 
to those shown in Fig.~\ref{CumulativeJstars}.}
\label{Convergence1}
\end{center}
\end{figure}

We perform a `strong' convergence\footnote{{Strong convergence test refers to comparing simulations with the same subgrid physics and parameters, 
as well as volume and initial conditions, but with different resolutions.}} test 
(see S15 for a discussion on `strong' and `weak' convergence)
of the resolution we use throughout this work (see Table~1).
To do this we use a smaller volume, but same resolution as the 
simulation described in Table~1, and a run with the same box size 
but higher resolution (see Table~\ref{TableSimus2} for the details of the 
simulations). \citet{Schaller15} and \citet{Lagos16b} have already presented detailed 
convergence tests for the mass and velocity radial distribution of galaxies, 
and angular momentum, respectively, in \eagle.
Here we focus on the radial profiles of $j_{\rm stars}$ of galaxies in \eagle\ that 
have (not) had mergers.

Fig.~\ref{Convergence1} shows the $j_{\rm stars}$ radial profiles of galaxies that 
have not gone through mergers ($N=0$), and those that had at least one merger ($N\ge 1$) by $z=0$ 
in \eagle. The difference between the $N=0$ and $N\ge 1$ is very similar 
in the two simulations despite their different in resolution. This shows 
that the profiles analysed in this work are well converged 
at the resolution adopted in the simulation of Table~1.

\section[]{Radial $j_{\rm stars}$ profiles at fine time intervals between outputs}\label{SnipshotTests}

\begin{figure}
\begin{center}
\includegraphics[trim=0mm 4mm 0mm 4mm, clip,width=0.4\textwidth]{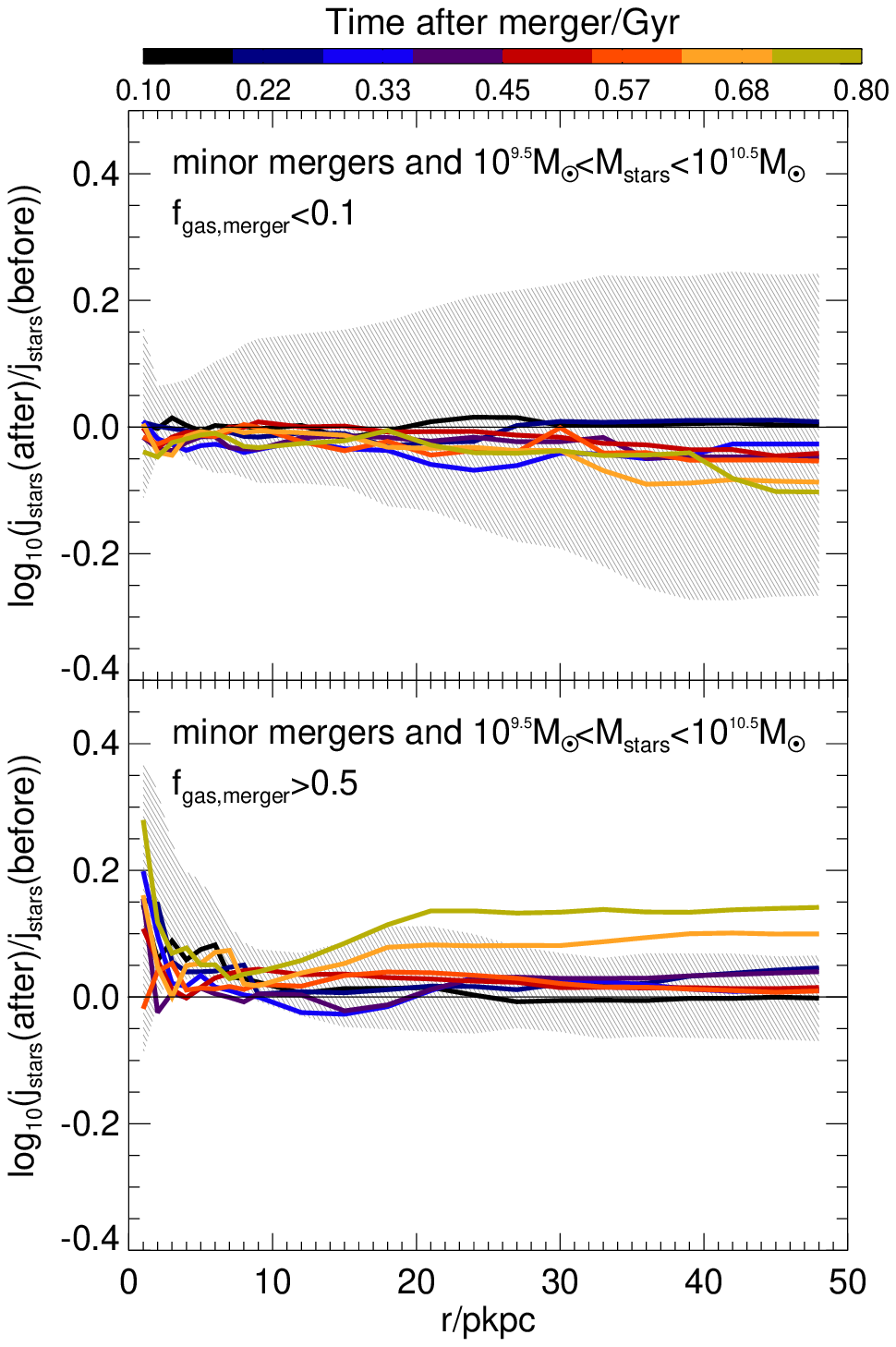}
\caption{The ratio between the mean $j_{\rm stars}$ after and before galaxy mergers, measured in an aperture $r$, as a function of $r$.
We measure $j_{\rm stars}$ after the merger in $8$ subsequent snipshots after the merger. Each snipshot 
samples a timestep of $\approx 0.1$~Gyr.
Here we show galaxies with $10^{9.5}\,\rm M_{\odot}<M_{\rm stars}<10^{10.5}\,\rm M_{\odot}$ 
that went through a minor merger in the redshift range $0.65\lesssim z\lesssim 0.75$. 
The top panel shows the subsample of gas-poor mergers, while the bottom panel shows gas-rich mergers.
Lines show the median, with the colour indicating the time after the merger, as shown in the colorbar at the top.
For simplicity we only show the $25^{\rm th}$-$75^{\rm th}$ percentile range (shaded region) for the first snipshot after the merger.
For reference, the horizontal lines show no change on $j_{\rm stars}(r)$, and so
 values above the line show an increase in $j_{\rm stars}$, while the opposite holds if below the line.}
\label{SnipshotMinorMergers}
\end{center}
\end{figure}

\begin{figure}
\begin{center}
\includegraphics[trim=0mm 4mm 0mm 4mm, clip,width=0.4\textwidth]{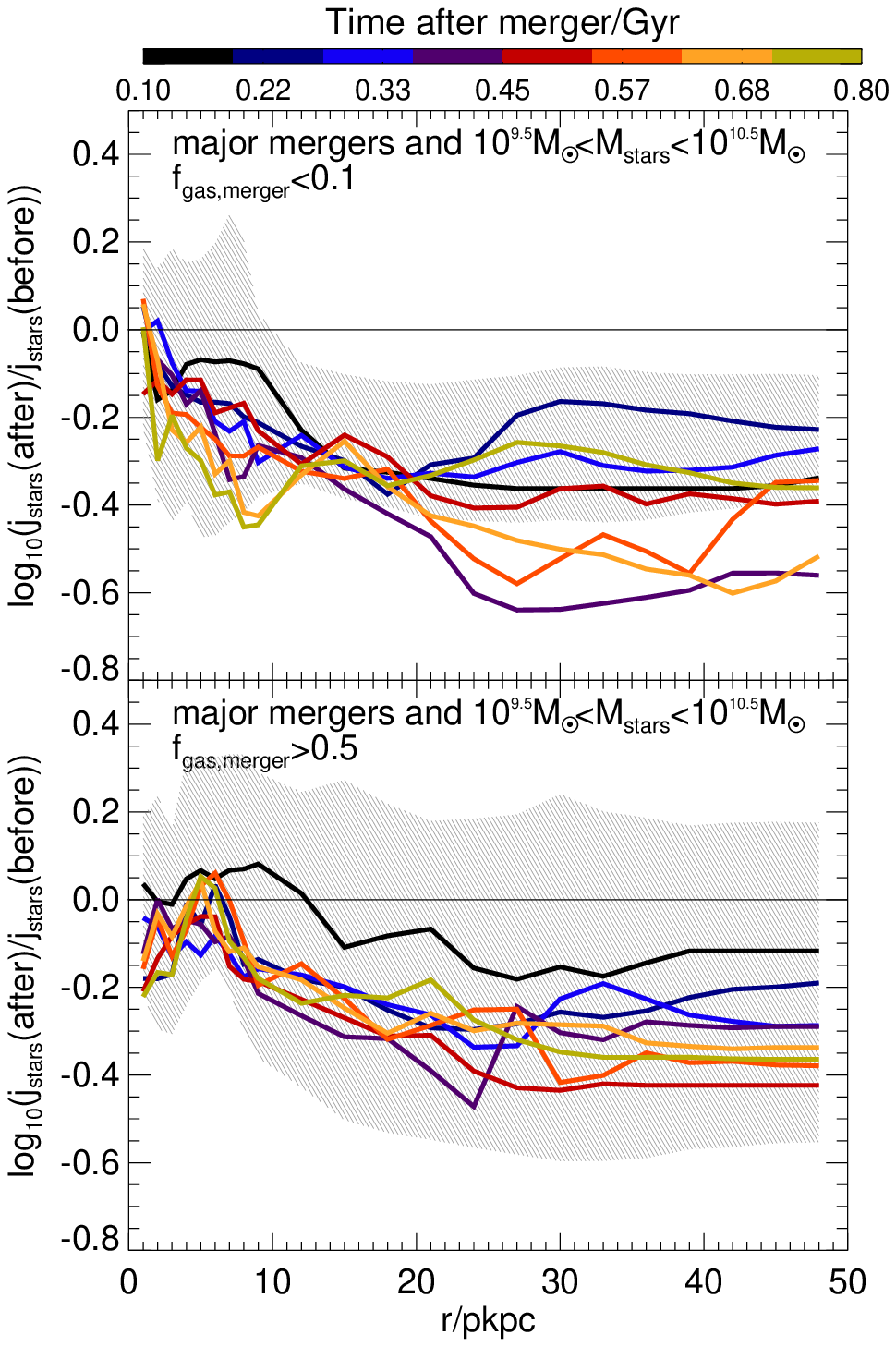}
\caption{As Fig.~\ref{SnipshotMinorMergers} but for major mergers.}
\label{SnipshotMajorMergers}
\end{center}
\end{figure}

The standard trees of \eagle\ connect $29$ epochs for which snapshots are saved (including all particle properties). 
The time span between snapshots can range from $\approx 0.3$~Gyr to $\approx 1$~Gyr.
Galaxy mergers, however, may require finner time intervals between snapshots to follow in more detail how the merger evolves.
\citet{Schaye14} showed than in addition to the snapshots described above, the \eagle\ simulations 
also record $400$ {\it snipshots}, in the redshift range $0\le z\le20$, saving fewer gas particle properties.
In our case, we would like to measure the mean radial $j_{\rm stars}$ profile in galaxies during and after the merger, 
and the information stored in snipshots allows us to do this. 
Owing to the computational expense of applying {\tt SUBFIND} to the outputs of \eagle\, only $200$ even-numbered snipshots 
of the simulation suite were catalogued. This decreases the time span between snipshots
to $\approx 0.05$~Gyr to $0.3$~Gyr. 

Here, we take all the snipshots between $z\approx 0.5$ and $z\approx 1$ and select all galaxy mergers 
that took place in that redshift range. We focus on this range because it is roughly when gas-rich and gas-poor mergers 
happen in similar numbers (see Fig.~\ref{FrequencyMergers}) in the Ref-L0100N1504 simulation. 
We calculate the radial $j_{\rm stars}$ profiles before and after the galaxy mergers (from $\approx 0.1$ to $\approx 0.8$~Gyr
after a minor merger, in timesteps of $\approx 0.1$~Gyr). 
We show in Fig~\ref{SnipshotMinorMergers} the radial $j_{\rm stars}$ profiles after the merger divided by the profiles 
before the merger for galaxies with $10^{9.5}\,\rm M_{\odot}<M_{\rm stars}<10^{10.5}\,\rm M_{\odot}$ 
We separate 
mergers into gas-rich and gas-poor. 
Our idea here is to test if the results of Fig.~\ref{DeltaJDeltaM_minormajorc} are affected by 
how fine the time interval between outputs is in the simulation. 
We find that gas-poor minor mergers systematically decrease $j_{\rm stars}$ 
over the entire radial range, while gas-rich minor mergers help increase 
$j_{\rm stars}$ in the central parts of galaxies, while changing only mildly 
$j_{\rm stars}$ at $r\gtrsim 10$~pkpc. Note that at later times ($\gtrsim 0.6$~Gyr after the merger)
$j_{\rm stars}$ in the outer regions starts increasing faster. We interpret this behaviour as resulting 
from continuing star formation, rather than due to the galaxy merger. 

Fig.~\ref{SnipshotMajorMergers} is as Fig.~\ref{SnipshotMinorMergers} but for major mergers.
Although the trends are noisy, there is a systematic effect of gas-poor major mergers to decrease $j_{\rm stars}$ 
over the entire radial range probed here. Gas-rich major mergers tend to preferentially reduce 
$j_{\rm stars}$ at $r\gtrsim 10$~kpc, while not affecting the inner regions of galaxies much. Although 
noisy, one could even argue that $j_{\rm stars}$ increases in the inner regions of galaxies as a result of 
a gas-rich major merger. We find that the results here are broadly consistent with those presented in the 
top panels of Fig.~\ref{DeltaJDeltaM_minormajorc}, and thus we conclude that finer time resolution only confirms the behaviour 
we analysed there.

\section[]{The effect of redshift, stellar mass and aperture in $j_{\rm stars}$}\label{ExtraTests}
\begin{figure*}
\begin{center}
\includegraphics[trim=0mm 4mm 0mm 1mm, clip,width=0.45\textwidth]{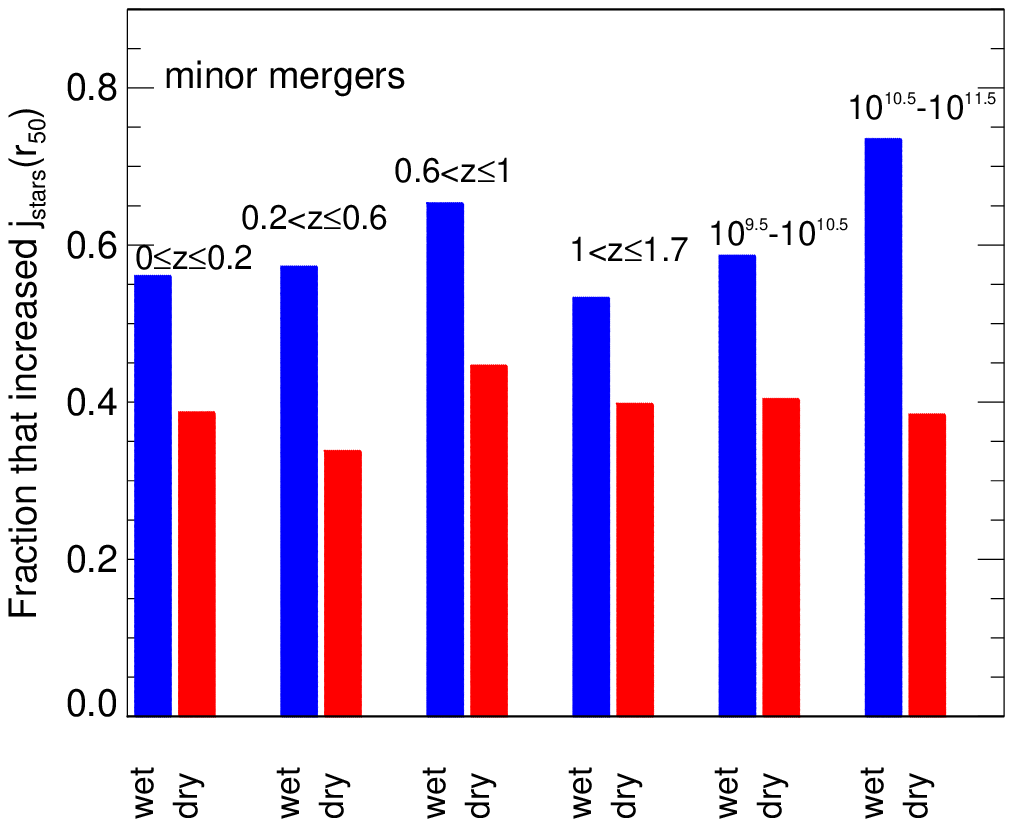}
\includegraphics[trim=0mm 4mm 0mm 1mm, clip,width=0.45\textwidth]{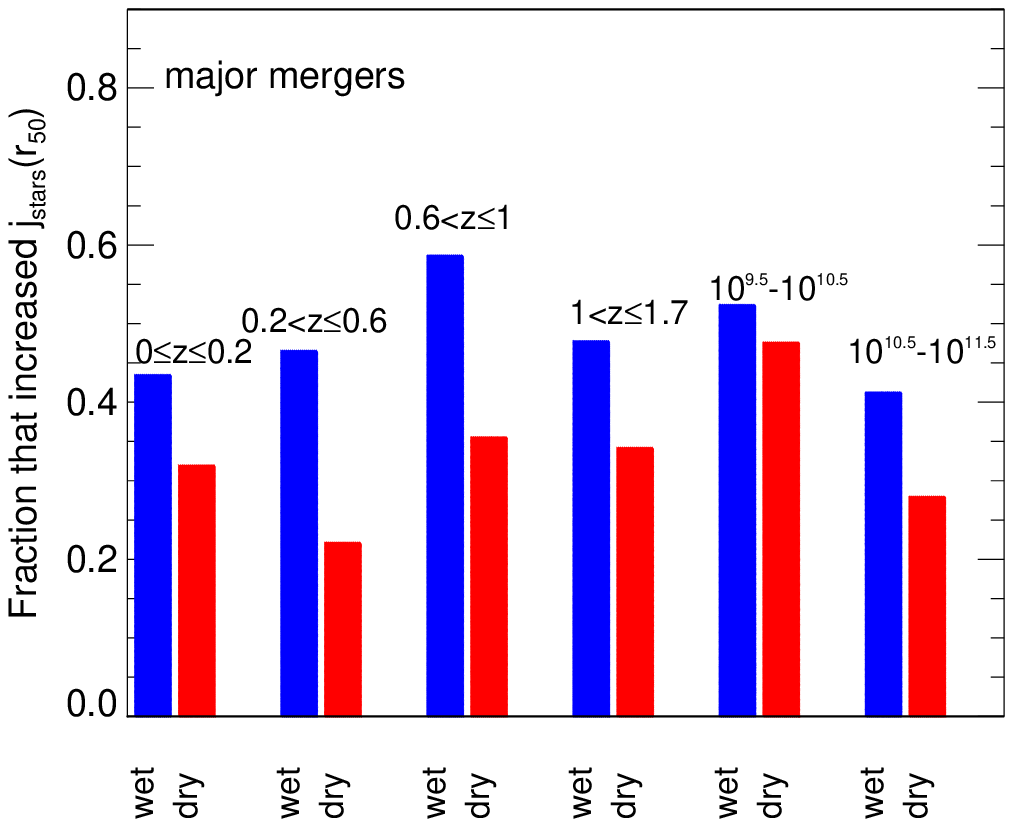}
\caption{{\it Left panel:} Fraction of galaxies that display an increase in their $j_{\rm stars}(r_{\rm 50})$ 
during minor mergers at different redshifts and in two bins of stellar mass, split into wet and dry mergers, as labelled.
{\it Right panel:} As in the left panel but for major mergers.}
\label{HistogramsDeltaJcMass}
\end{center}
\end{figure*}

\begin{figure*}
\begin{center}
\includegraphics[trim=0mm 24.5mm 0mm 1mm, clip,width=0.89\textwidth]{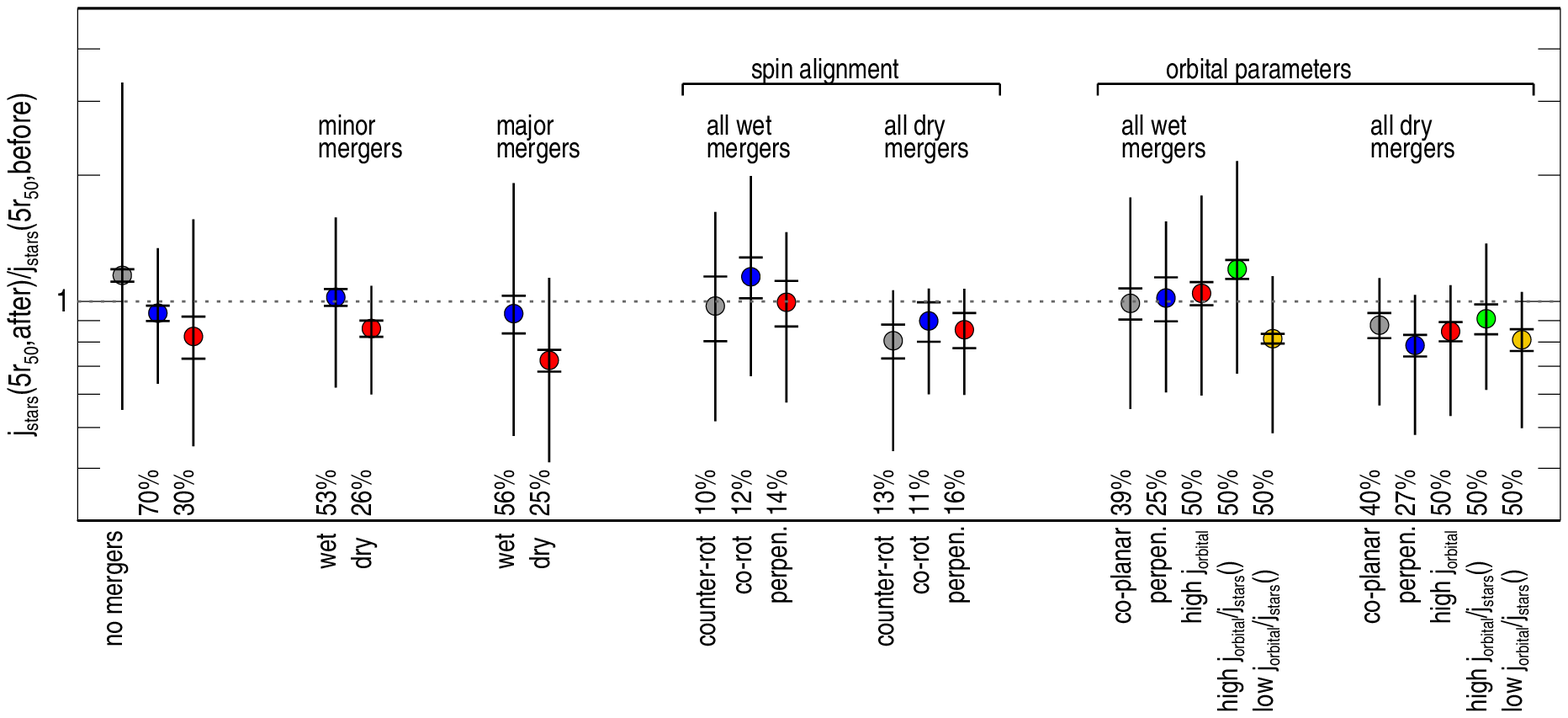}
\includegraphics[trim=0mm 0mm 0mm 1mm, clip,width=0.89\textwidth]{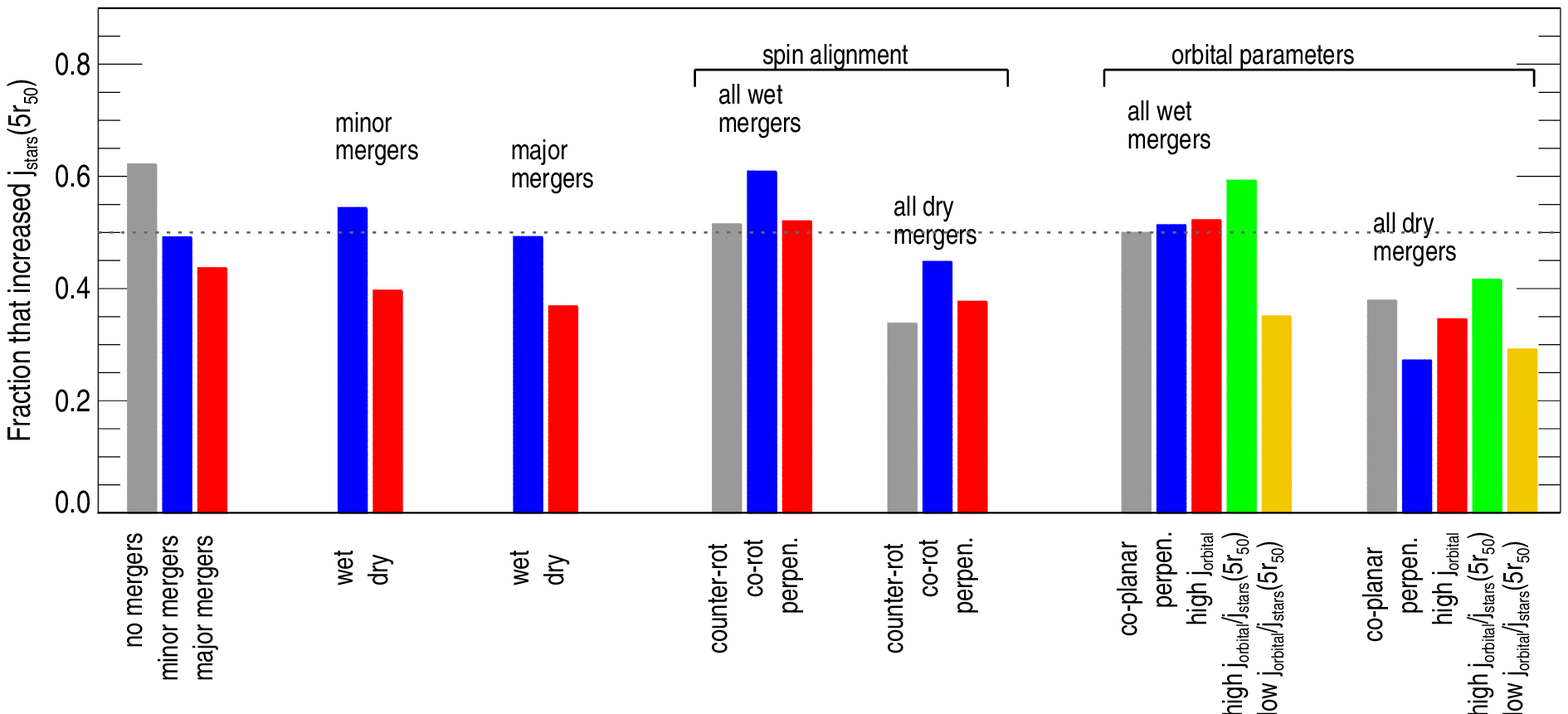}
\caption{As Fig.~\ref{HistogramsDeltaJ} but for $j_{\rm stars}(5r_{\rm 50})$.}
\label{HistogramsDeltaJ5}
\end{center}
\end{figure*}

In $\S$~\ref{temporalevo}, we stacked all of the galaxy mergers that take place in galaxies with $M_{\rm stars}\ge  10^{9.5}\,\rm M_{\odot}$
and in the redshift range $0\le z\le 2.5$. This may introduce significant biases due to the time-stepping of the
simulation (different snapshots cover different timescales), and also due to galaxies having very different sizes at different cosmic epochs.
In order to quantify that bias, we analyse galaxy mergers at different cosmic epochs and stellar mass bins, separated into minor and major and into wet and dry mergers 
in Fig.~\ref{HistogramsDeltaJcMass}. We first compare the distributions as a function of gas-richness, and we find that there is
no statistical difference between the wet and dry minor merger populations at different redshifts. The Kolmogorov-Smirnov p-values                  
are in the range $\approx 0.2-0.9$ when we compare wet or dry merger populations at different redshifts. 
When we compare wet vs. dry minor mergers at different redshift, we find that the differences seen in
Fig.~\ref{HistogramsDeltaJ} are always present with high statistical significance ($p$-values are $\lesssim 10^{-4}$).
When we analyse different stellar mass bins we reach the same conclusion.
Thus, we can comfortably assume that stacking minor mergers at different redshift does not introduce any significant bias to our analysis.
In the case of major mergers, we see more variations between the subsamples at different redshifts and stellar masses, but the difference between 
dry and wet mergers is still the most important one statistically (with $p$-values $\lesssim 10^{-3}$).

Fig.~\ref{HistogramsDeltaJ5} shows the ratio between $j_{\rm stars}(5\,\rm r_{50})$ after and before mergers (top panel) and the frequency in which 
mergers increase $j_{\rm stars}(5\,\rm r_{50})$ (bottom panel). We find that the results shown here are similar to those 
of Fig.~\ref{HistogramsDeltaJ} for $j_{\rm stars}$ measured within $r_{50}$.

\end{document}